\documentclass[9pt,twocolumn,twoside]{pnas-new}

\templatetype{pnasresearcharticle} 

\usepackage{enumitem}
\usepackage{booktabs}
\usepackage{tikz}
\usepackage{adjustbox}
\usepackage{multirow}
\usepackage{graphicx}
\usepackage{subfigure}
\usepackage{xcolor}
\usepackage{array}
\usepackage{kotex}
\usepackage{amsmath}
\usepackage{subcaption}

\definecolor{supportcolor}{RGB}{160,196,225}
\definecolor{opposecolor}{RGB}{242,161,165}   
\definecolor{amendcolor}{RGB}{246,211,178}    
\definecolor{monitorcolor}{RGB}{183,230,181}  

\definecolor{violet-5}{RGB}{132, 94, 247}

\begin{document}

\title{Measuring Interest Group Positions on Legislation: An AI-Driven Analysis of Lobbying Reports}


\author[a]{Jiseon Kim}
\author[a]{Dongkwan Kim}
\author[b]{Joohye Jeong}
\author[a, 1, *]{Alice Oh}
\author[b, 1, *]{In Song Kim}

\affil[a]{School of Computing, Korea Advanced Institute of Science and Technology (KAIST), Daejeon, Republic of Korea 34141}
\affil[b]{Political Science, Massachusetts Institute of Technology (MIT), Cambridge, MA 02142}

\leadauthor{Kim}

\significancestatement{


This study provides the first large-scale measurement of how special interest groups (SIGs) influence U.S. legislation by examining their positions on bills from the 111th to the 117th Congresses. Using advanced AI methods, such as large language models (LLMs) and graph neural networks (GNNs), we measure latent SIG positions on a wide range of bills. Notable patterns emerge from our analysis, including the relationship between SIG positions and legislative progress in the U.S., the impact of firm size on lobbying behavior, and variations in lobbying strategies across different bill topics and industries.
We release the bill position dataset and lobbying position scores (LPscores) to support future research on interest group preferences.
}

\authorcontributions{Author contributions: J.K., D.K., J.J., A.O., and I.K. designed research; J.K., D.K., and J.J. performed research; J.K., D.K., and J.J. analyzed data; and J.K., D.K., J.J., A.O., and I.K. wrote the paper.}
\authordeclaration{The authors declare no competing interest.}
\correspondingauthor{\textsuperscript{1}To whom correspondence should be addressed. E-mail: alice.oh@kaist.edu and insong@mit.edu}


\begin{abstract}

Special interest groups (SIGs) in the U.S. participate in a range of political activities, such as lobbying and making campaign donations, to influence policy decisions in the legislative and executive branches. The competing interests of these SIGs have profound implications for global issues such as international trade policies, immigration, climate change, and global health challenges. Despite the significance of understanding SIGs' policy positions, empirical challenges in observing them have often led researchers to rely on indirect measurements or focus on a select few SIGs that publicly support or oppose a limited range of legislation. This study introduces the first large-scale effort to directly measure and predict a wide range of bill positions---\textit{Support}, \textit{Oppose}, \textit{Engage} (\textit{Amend} and \textit{Monitor})---across all legislative bills introduced from the 111th to the 117th Congresses. We leverage an advanced AI framework, including large language models (LLMs) and graph neural networks (GNNs), to develop a scalable pipeline that automatically extracts these positions from lobbying activities, resulting in a dataset of 42k bills annotated with 279k bill positions of 12k SIGs. 
With this large-scale dataset, we reveal (i) a strong correlation between a bill’s progression through legislative process stages and the positions taken by interest groups, (ii) a significant relationship between firm size and lobbying positions, (iii) notable distinctions in lobbying position distribution based on bill subject, and (iv) heterogeneity in the distribution of policy preferences across industries. We introduce a novel framework for examining lobbying strategies and offer opportunities to explore how interest groups shape the political landscape.
\end{abstract}

\dates{This manuscript was compiled on \today}
\doi{\url{www.pnas.org/cgi/doi/10.1073/pnas.XXXXXXXXXX}}

\maketitle
\thispagestyle{firststyle}
\ifthenelse{\boolean{shortarticle}}{\ifthenelse{\boolean{singlecolumn}}{\abscontentformatted}{\abscontent}}{}

\firstpage[14]{4}


Who makes the laws? In the U.S., while elected officials formally craft legislation, the process is heavily shaped by other powerful actors—most notably special interest groups (SIGs) such as corporations, labor unions, and issue advocacy groups~\citep{baumgartner2009lobbying, grossman:helpman:01}. Interest groups\footnote{In this paper, we refer to special interest groups(SIGs) simply as interest groups.} engage in lobbying to influence or subsidize policymakers’ positions on a wide range of issues~\citep{hall:dear:06}, and their competing interests play a central role in shaping legislation on global challenges including international trade~\citep{kim:17}, climate change~\citep{farrell2016pnas, brulle2014institutionalizing, stokes2020short, xie2025tracing}, economic inequality~\citep{melissa2017pnas, dietze2021framing, wang2024global}, foreign policy~\citep{milner:ting:15}, and immigration~\citep{liao:2023}. Thus, examining the policy preferences of individual interest groups, especially when they lobby specific legislative bills, offers critical insights into their influence on legislative outcomes.

Despite significant progress in research enabled by the Lobbying Disclosure Act of 1995, which helps scholars identify which interest groups lobby on which legislative bills~\citep{kim2021mapping}, major empirical challenges persist in observing the \textit{positions} interest groups take, such as whether they support or oppose the bills. Consequently, scholars often limit their analyses to selected interest groups within specific states~\citep{grasse2011influence, daniel2022prq} or focus solely on particular policy issues~\citep{kim:17}, resulting in a lack of broader coverage.

Other studies seeking to systematically measure policy preferences across a wide range of interest groups also face several limitations. First, research relying on indirect measures to infer interest group positions may not accurately reflect the actual stances that groups express during the legislative process. For example, the Database on Ideology, Money in Politics, and Elections (DIME)~\citep{bonica2014mapping} provides campaign finance scores (CFscores) based solely on donation patterns to politicians whose ideological positions are estimated from sparse roll-call votes~\citep{bonica2019donation}. This method assumes donations align closely with candidates' ideologies, even though interest groups may not explicitly take policy positions. Second, datasets compiled by activists and nonprofit organizations, such as MapLight~\citep{lorenz2020large}, manually code interest group positions by examining publicly available sources (e.g., company websites, press releases, government records). While valuable, these datasets generally capture revealed preferences only on selected legislation. Moreover, as discussed further below, many interest groups in these datasets do not actively engage in lobbying, providing limited and indirect insights into their broader impacts on the legislative process. Third, interest group lobbying encompasses more nuanced actions beyond simply supporting or opposing legislation, including activities like proposing amendments, continuous monitoring, and incremental policy adjustments~\citep{butl:mill:22}. Finally, manual data collection methods in prior studies are costly, difficult to scale, and challenging to reproduce, limiting their application to large-scale analyses.

\begin{figure*}[t]
\centering
\includegraphics[width=\linewidth]
{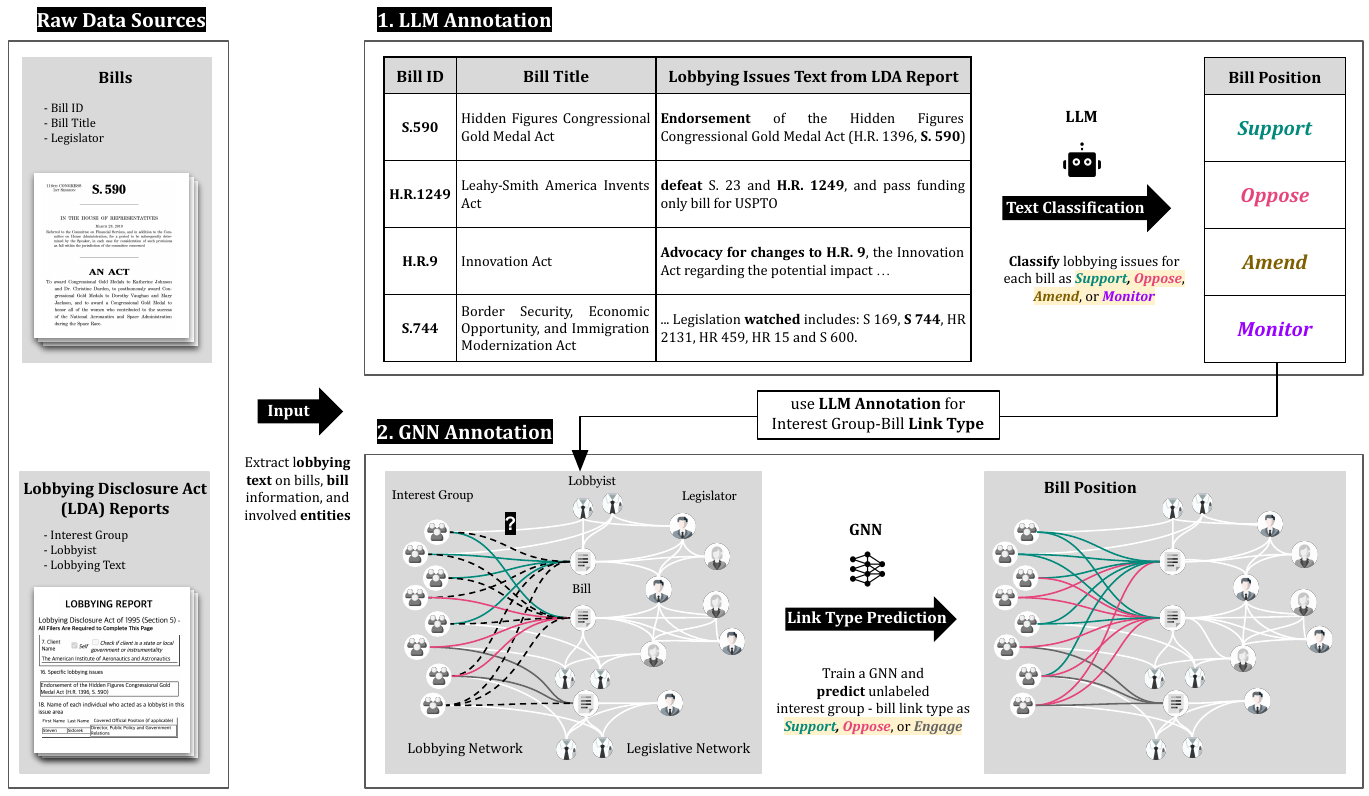}
\caption{Overview of bill position annotation pipeline. This  outlines the two approaches used in our bill position annotation process. First, in \textit{LLM Annotation} stage, text information from raw data sources is extracted, and positions are annotated using a large language model (LLM). When the text alone does not offer enough clarity to determine the bill position, lobbying and legislative network data are incorporated in the \textit{GNN Annotation} pipeline, where a graph neural network (GNN) is used to refine the prediction. By minimizing human annotation, the pipeline is made more scalable, while still capturing the bill positions of interest groups involved in the legislative process, providing a dataset that reflects the real-world legislative activities and a wider range of lobbying types.}
\label{fig:overview}
\end{figure*}

In this paper, we address these limitations by introducing the first large-scale measurement of interest group positions on individual bills they have directly lobbied from the 111th to 117th U.S. Congresses. Unlike previous studies, our approach captures the explicit positions interest groups take during the lobbying process---not only support and opposition but also more nuanced stances such as amending or monitoring legislation. To ensure scalability and reproducibility, we develop an automated annotation pipeline that combines Large Language Models (LLMs) to systematically analyze lobbying reports submitted by individual interest groups with Graph Neural Networks (GNNs) to model the structure of lobbying and legislative networks. Specifically, we identify systematic patterns associated with different types of political position-taking, as reflected in the high-dimensional network connections among interest groups, lobbyists, and policymakers, along with their observable characteristics. Applying this pipeline, we provide a comprehensive measurement that uncovers previously hidden policy preferences of interest groups in the legislative process. Our measurements serve as a foundational resource for analyzing legislative processes, enabling the exploration of key dimensions such as bill outcomes, firm characteristics, policy topics, and industry involvement. The primary contributions of our study are as follows:

\vspace{-5pt}
\begin{itemize}
    \item \textbf{Comprehensive Bill Position Dataset}: We construct a dataset that closely mirrors real-world legislative activity by incorporating multiple political actors and capturing a range of bill positions---\textit{Support}, \textit{Oppose}, and \textit{Engage} (i.e., \textit{Amend} and \textit{Monitor})---to offer a more detailed view of lobbying behavior.
    \item \textbf{Scalable Bill Position Annotation Pipeline}: We introduce an automated, scalable pipeline for generating bill position data by combining Large Language Models (LLMs) to analyze lobbying reports with Graph Neural Networks (GNNs) to model lobbying and legislative networks.
    \item \textbf{Analysis of Lobbying Strategies and Hidden Interests}: Our analysis reveals distinct patterns in lobbying strategies and uncovers the latent interests of interest groups, showing how these relate to bill outcomes, firm characteristics, policy topics, and industry sectors.
\end{itemize}
We publicly release the bill position dataset and lobbying position scores (LPscores) to support research on interest group politics and computational social science.

\section*{Bill Position Annotation Pipeline}

We introduce a scalable method for constructing a bill position dataset. Fig.~\ref{fig:overview} provides an overview of our pipeline. Starting from raw data sources, including \textit{Bills} and \textit{Lobbying Disclosure Act (LDA) reports}, our annotation process follows a two-step approach. First, we employ a large language model (LLM) to extract bill positions based on bill IDs, titles, and raw lobbying descriptions. Then, we apply a graph neural network (GNN) on lobbying and legislative network data with LLM-generated annotations. This step helps infer bill positions more accurately by leveraging network connections, especially when lobbying reports alone are unclear for annotation.

As a result, using our pipeline with LLM and GNN, we obtain a total of 279,104 bill position data across 12,032 interest groups and 42,475 bills as illustrated in Table~\ref{tab:bill_position_dataset_size}.

\matmethods{
\paragraph{Raw Data Source}
Our study covers 14 years of lobbying activities in the legislative process from the 111th to the 117th Congress, from 2009 to 2022.\footnote{The 111th Congress marks the point where lobbying data became more structured and transparent following the full implementation of HLOGA (2007) and stricter disclosure rules under the Obama administration. See HLOGA of 2007, Pub. L. No. 110-81, 121 Stat. 735 (2007); Executive Order 13490, January 21, 2009.} Lobbying reports, which are submitted quarterly by lobbyists on behalf of their clients, detail their lobbying efforts, the firms involved, the participating lobbyists, and the nature of the lobbying activities. We collected all lobbying reports filed during this period from the Senate Office of Public Records. We linked these lobbying activities to various bill-level details such as the bill's title, the sponsoring and co-sponsoring legislators, and its status throughout the legislative process. We gathered this data from \href{https://www.congress.gov/}{Congress.gov}, the primary government source.

\subsection*{1. LLM Annotation} 
We employ an LLM to extract bill positions from lobbying activity descriptions, effectively leveraging textual information in the source data.

\paragraph{Preprocessing} 
In the raw lobbying reports, interest group activities are documented in Section 16 under `\textit{Specific Lobbying Issues}.' These descriptions often encompass multiple lobbying activities and can be quite extensive. To improve the efficiency of LLM annotation, we segment them into individual lines. It is crucial to note that interest groups are legally mandated to report any congressional bills they have lobbied on, as required by the Lobbying Disclosure Act of 1995.  To identify lobbying activities related to specific bills, we select descriptions that reference bill IDs from the entire set of lobbying texts.
The types of bills considered in this study are limited to those that are formally enacted as law, specifically Senate and House bills.
From this pool of approximately 1.4 million bill-line text pairs, we further refine the dataset by selecting entries that contain keywords indicative of bill positions, such as \textit{`endorse', `defeat', `advocate'}, and \textit{`watch.'}
The keyword list is structured based on the definitions of lobbying and activity provided by each state,\footnote{These definitions include terms related to lobbying and lobbyists, as outlined in the statutes of the respective states. See \url{www.sos.state.co.us/pubs/lobby/files/guidanceManual.pdf}} as well as lobbying guidance documents issued by state authorities.\footnote{For reference, we use the \textit{Colorado Secretary of State Lobbying Guidance Manual}. See \url{www.ncsl.org/ethics/how-states-define-lobbying-and-lobbyist}} A comprehensive list of keywords is available in SI Appendix~\ref{sup:llm_annotation}. An example of a lobbying description can be found in Fig.~\ref{fig:overview}, under the column labeled \textit{`Lobbying Issues Text from LDA Report.'}

\paragraph{Text Classification}
In our approach, we use Azure OpenAI to perform labeling with GPT-4 (\texttt{2023-09-01-preview}), setting the temperature to 0 to ensure deterministic outputs. As shown in Fig.~\ref{fig:overview} at ``\textit{1. LLM Annotation} stage'', we provide the LLM with instructions for bill position classification, supplying the target bill ID along with bill-related information, such as the bill title. The LLM then generates an output with one label based on the given prompt, classifying the lobbying description associated with the bill into a specific position. (The full text input examples and prompt are described in SI Appendix~\ref{sup:llm_annotation}.) If there is insufficient information to definitively classify the lobbying description into a particular bill position, we categorize it separately under \textit{`Mention.'} This ensures that ambiguous cases are distinctly labeled, preventing misclassifications due to a lack of context.

\paragraph{Validation}
To evaluate the quality of LLM annotations on bill position classification, we examined the four categories defined in the bill annotation dataset: \textit{Support}, \textit{Oppose}, \textit{Amend}, and \textit{Monitor}. Two authors independently labeled the data, and we selected 391 samples for which both assigned the same label, reflecting clear expressions of bill position. The comparison between human labels and GPT-4's annotations shows a high alignment, with accuracy and F1 scores of 96.93 and 97.19, respectively.
Additionally, we verify the alignment of GPT-4's annotations with the existing MapLight bill position dataset.\footnote{\url{https://www.maplight.org/data-series}} 
Our dataset shares 1,182 overlapping interest group–bill ID pairs with the MapLight data.\footnote{Since MapLight provides only support and oppose labels, we limit our comparison to these categories.} Among these, 1,076 pairs (91.03\%) have matching bill positions, indicating a strong alignment despite differences in data sources.

\begin{table}[]
\centering
\caption{Size of bill position dataset 
}
\label{tab:bill_position_dataset_size}
\begin{tabular}{@{}crrr@{}}
  & LLM Annotation & GNN Annotation           & Overall                  \\ \midrule
Interest Group & 6,495          & 11,185                   & 12,032                   \\
Bill           & 22,820         & 36,824                   & 42,475                   \\ \midrule
Support        & 25,099         & 54,204                   & 79,303                   \\
Oppose         & 5,706          & 5,456                    & 11,162                   \\
Amend          & 9,317          & \multirow{2}{*}{137,023} & \multirow{2}{*}{188,639} \\
Monitor        & 42,299         &                          &                          \\ \midrule
Total          & 82,421         & 196,683                  & 279,104                  \\ \bottomrule
\end{tabular}
\end{table}



\subsection*{2. GNN Annotation}
When textual information may be limited in conveying the bill position—such as in cases where lobbying text lacks key indicators or is classified as \textit{`Mention'}—we enhance our classification by incorporating network information that emerges during the legislative process. Prior research~\citep{abi2023ideologies, kim2021mapping} has shown that legislative network structures play a crucial role in predicting political activities, underscoring their relevance in this context. By leveraging these network-based insights, we complement textual analysis and improve the accuracy of bill position inference.

\paragraph{Graph Configuration}
The ``\textit{2. GNN Annotation} stage'' section depicted in the lower part of Fig.~\ref{fig:overview} illustrates the pipeline for annotating interest groups' bill positions using high-dimensional policy network data. This graph includes various entities involved in lobbying activities and the legislative process, such as interest groups, bills, legislators, and lobbyists.
In the lobbying network, connections are derived from lobbying reports, linking interest groups, lobbyists, and bills based on their collaborative advocacy efforts. For instance, when an interest group hires lobbyists to support a bill, a connection forms between them and the bill.
The legislative network captures relationships between bills and legislators, where links are established through sponsorship or cosponsorship. Additionally, a connection between a lobbyist and a legislator is formed if they have previously worked together.\footnote{We gather this information from ``Covered Official Position'' reported under Section 18 of lobbying reports.} One of the primary contributions of our study is leveraging this high-dimensional network structure to infer systemic patterns associated with latent positions, drawing on our understanding of other political actors' preferences. Put simply, if actors A and B consistently lobby in similar ways within comparable political networks and A's position toward a specific bill is known, we can reliably predict B’s position on the bill, even if it hasn’t been explicitly disclosed.

We implement a heterogeneous GNN for bill position annotation since interest groups, bills, and lobbyists represent distinct entity types. Each node in the graph is represented by an embedding that encodes its relevant attributes. Interest groups, for example, are described by attributes such as state, industry code, and government affiliation. Bills include details like their subject, the sponsoring party, and their final status in the legislative process. Likewise, lobbyists are characterized by features including ethnicity, gender, and past party affiliation for former Congress members.
For more details on additional graph configurations and feature definitions for other entities, refer to the SI Appendix~\ref{sup:gnn_annotation}. 

During the initial experiments, we found that classifying bill positions into four categories, including \textit{Amend}, led to significantly lower accuracy for this particular class. To improve model performance and ensure more reliable results, we decided to merge \textit{Amend} and \textit{Monitor} into a single \textit{Engage} category. As a result, the task was reformulated as a 3-class classification problem: \textit{Support}, \textit{Oppose}, and \textit{Engage}. This consolidation allowed the model to focus more effectively on the \textit{Support} and \textit{Oppose} categories, leading to more accurate and meaningful classifications overall.

\paragraph{Link Type Prediction}
The task of the GNN is to predict the type of link between interest groups and bills (i.e., the bill position of interest groups). GNN is trained and evaluated using LLM annotations and labels from the MapLight dataset.\footnote{The sample from MapLight, mapped to the lobbying pairs organized in LobbyView, includes 18,180 \textit{Support} and 4,455 \textit{Oppose} positions, totaling 22,635 bill position pairs used for GNN training.} 
The bill position dataset is divided into train, validation, and test sets in a ratio of 7:1:2.
We conduct a hyperparameter search to determine the optimal GNN configuration and provide details on the search space and settings in the SI Appendix~\ref{sup:gnn_annotation}.

The model with the best configuration achieves a solid performance with an overall accuracy of 78.51 and an F1 score of 74.65.\footnote{These scores represent the mean and standard deviation of predictions made five times with different random seeds.} The F1 scores for each class are as follows: \textit{Support} at 79.19, \textit{Oppose} at 64.06, and \textit{Engage} at 80.70, all of which outperform the baseline configurations.
Using the trained model, we predict the bill positions for unlabeled edges between interest groups and bills, treating these predictions as large-scale proxy labels. Bill positions are selected from the model's predictions when the class probability is higher than 0.9.\footnote{Youden’s Index is a measure that combines sensitivity (correctly identifying positives) and specificity (correctly identifying negatives) to find the best cutoff for model predictions. In our model, the Youden’s Index suggests a minimum cutoff of around 0.58. However, we use a threshold of 0.9 to ensure more reliable results, minimizing false positives and improving the model's overall trustworthiness.}
As a result, through GNN annotation, a total of 196,683 bill position datasets are generated, as shown in Table~\ref{tab:bill_position_dataset_size}.

\subsection*{3. Bill Position Dataset Overview and Comparison}
\paragraph{Overview of Our Dataset}
Our dataset covers bill positions from the 111th to the 117th Congress, providing a comprehensive and scalable approach for capturing the stance of interest groups on various legislative matters. The dataset includes 279,104 bill positions between 12,032 interest groups and 42,475 bills, ensuring broad coverage across a wide range of legislative activities, as shown in Table~\ref{tab:bill_position_dataset_size}.\footnote{We exclude any instances where the interest group’s bill position is assigned multiple labels or conflicts with labels from sources such as LLM, GNN, or MapLight, ensuring that each interest group has only one bill position per bill. The analysis that follows is based on this cleaned dataset.} To the best of our knowledge, this is the first large-scale measurement of interest group positions that actively engage in lobbying activities.

By comparing our dataset with the existing MapLight bill position dataset, we highlight its broader coverage and scale. Furthermore, we explain the differences between latent preferences based on our lobbying activity data and those based on campaign finance data, a widely used metric provided by DIME.

\paragraph{Comparison of Data Coverage with the MapLight Dataset}
We compare the scale of our dataset with the coverage of the existing human-curated MapLight bill position dataset,\footnote{This comparison is made with our original dataset, before excluding instances that conflict with MapLight bill position labels.} which spans the 111th to 115th Congresses. In contrast, our dataset covers the 111th to 117th Congresses, enabling direct comparisons of our measures against human-labeled interest group positions. MapLight lists a total of 194,479 interest group–bill position pairs; however, after mapping these pairs from lobbying reports,\footnote{We used Google Search to map each interest group name. The first relevant result—typically the group’s homepage or official profile—was selected. Ambiguous or generic results were manually reviewed for confirmation.} only 25,382 pairs involving 2,195 interest groups and 5,780 bills remain. The fact that only 13.1\% of the MapLight dataset matches our bill position data suggests that most MapLight interest groups do not actively lobby, providing limited insight into their broader legislative impact through direct lobbying activities. Moreover, our measures encompass over five times more interest groups (12,032) and seven times more bills (42,475). Thus, our proposed pipeline provides a scalable framework that can be directly applied to future congressional data, significantly enhancing coverage while reducing the need for human annotation.

\paragraph{Lobbying position scores (LPscores)}
Similar to how roll call vote records are used to estimate legislators' ideal points~\citep{poole1985spatial, clinton2004statistical}, we construct a one-dimensional latent preference embedding for interest groups based on their bill positions. We refer to these as Lobbying Position Scores (LPscores). By aggregating LLM and GNN annotations along with the MapLight dataset, we generate latent interest scores using an item response theory (IRT) model to provide a unified measure of their positions. To estimate these scores, we apply the graded model~\citep{samejima1968estimation} within the MIRT (Multidimensional Item Response Theory) package~\citep{chalmers2012mirt} in R, which allows us to handle ordered categorical labels. We focus on \textit{Support} and \textit{Oppose} labels, excluding \textit{Engage} to maintain consistency with established approaches in political analysis and to better capture distinct policy preferences. To ensure reliable estimation, we retain only interest groups that have lobbied on at least 10 bills and bills that have been lobbied by at least 10 different groups. After this filtering, we derive latent preferences for 1,796 interest groups and 2,020 bills. We evaluated the reliability of the MIRT results and found an overall Expected A Posteriori (EAP) reliability of 0.95, indicating that the estimated trait scores are highly stable and precise. The standard errors ranged from 0.03 to 0.30, all below the commonly accepted threshold of 0.32, confirming that the estimated scores are sufficiently reliable.

\paragraph{Comparison of Interest Group Preferences: CFscores vs. LPscores}
Campaign finance scores (CFscores), provided by DIME~\citep{bonica2023database}, are widely used to represent interest groups' ideological positions based on campaign finance data. To compare these with our LPscores, which provide a direct measure of lobbying activity, we examine 218 interest groups with identical names in both datasets.
Since CFscores are reported for individual Congresses, we first compute each group's mean CFscore across all Congresses and compare it to their bill position-based latent preference. The Pearson correlation between the two is 0.191,\footnote{Pearson correlation coefficient was computed using SciPy library. See \url{https://docs.scipy.org/doc/scipy/reference/generated/scipy.stats.pearsonr.html}} suggesting a positive but weak relationship. This indicates that while campaign contributions and lobbying activities may be related, they capture distinct aspects of interest group behavior.

\paragraph{Data Availability}
The replication codes for LLM, GNN annotation, and analysis are available at \url{https://github.com/hikoseon12/lobbying-position}, and the input and output data used for replication are available at the Harvard Dataverse: \url{https://doi.org/10.7910/DVN/D0QWM2}.
}
\showmatmethods{}

\section*{Analysis}
Our bill position dataset serves as a foundational tool for understanding the underlying lobbying positions and strategies of interest groups and enables a variety of analyses. We illustrate this with four example findings from the following analyses. (1) We first examine the timeline of interest groups' bill positions, tracking positions across different stages of a bill’s progress. 
(2) Secondly, we investigate how corporate lobbying is related to firm size.
(3) We then investigate how the subject matter of the bills leads to variations in the distribution of lobbying positions, uncovering distinct preferences across different bill topics. 
(4) Finally, we explore the variation in interest group preferences across industries, offering insights into how they shape lobbying behavior and influence policy directions.

\subsection*{Analysis 1: At What Stages Do Interest Groups Lobby with Bill Positions}
Building on previous research by \citep{Holyoke02102019, holyoke2022strategic}, which shows how lobbying activities vary as a bill moves through different stages of the legislative process, our study goes a step further by analyzing the quarterly lobbying reports to estimate when lobbying happens. We map interest groups' positions (\textit{Support}, \textit{Oppose}, \textit{Amend}, and \textit{Monitor}) to specific bill stages. This allows us to get a detailed view of how lobbying intensity increases as bills approach enactment, revealing the strategic nature of lobbying during key legislative junctures.
Since a bill can be lobbied at various points during its legislative journey, we estimate the timing of lobbying based on when a lobbying report is submitted. Given that reports are filed quarterly, we treat the submission deadline as a proxy for the timing of lobbying, assuming that the lobbying efforts reported were directed at the bill's most recent legislative stage.\footnote{We follow the quarterly deadline for lobbying reports as specified by Congress. For more details, see \url{https://lda.congress.gov/ld/help/default.htm?turl=Documents/LD2Requirements.htm}.} We then analyze how the four classes of bill position are distributed across the bill's stages using the LLM annotation.

\begin{figure}[t]
\centering
\includegraphics[width=\linewidth]{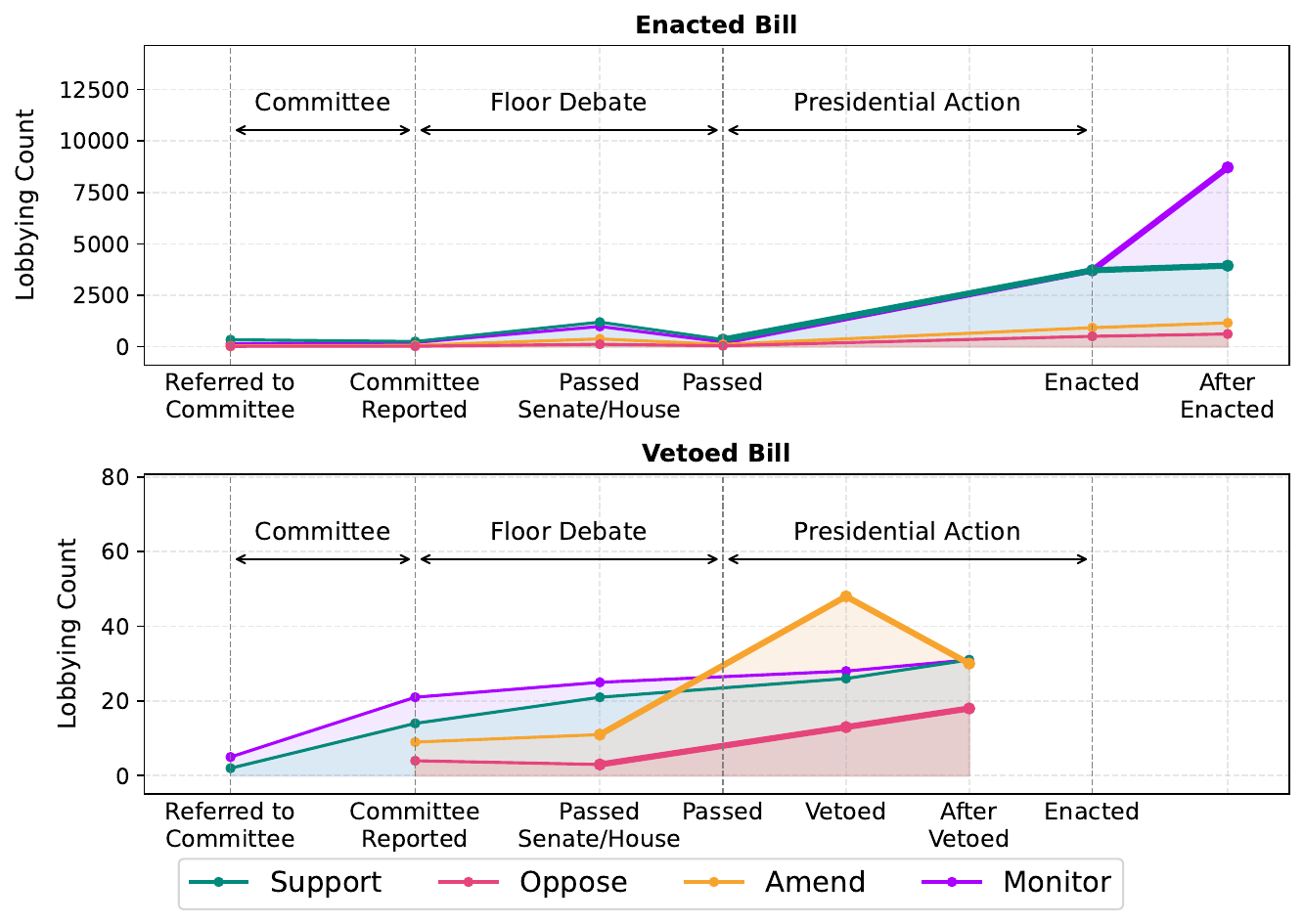}
\caption{Frequency of interest group positions across different legislative stages. We compare the bill position distributions of interest groups across different legislative stages. The x-axis represents the stages of the legislative process, from bill introduction to enactment, with lines depicting the frequency of positions: \textit{Support} (blue), \textit{Oppose} (red), \textit{Amend} (yellow), and \textit{Monitor} (green). This analysis includes bills that were ultimately `Enacted' and `Vetoed'. Bills with a higher frequency of \textit{Support} are more likely to advance through the legislative process (`Enacted'), while those with significant increases in \textit{Oppose} or \textit{Amend} typically fail to progress (`Vetoed'). We observe a clear correlation between interest group positions and the eventual legislative outcome of the bills.}
\label{fig:bill_position_stage}
\end{figure}

Fig.~\ref{fig:bill_position_stage} shows the distribution of bill positions across different stages in the legislative process, from the bill’s introduction to its final outcome. The colors of the lines represent: green for \textit{Support}, pink for \textit{Oppose}, yellow for \textit{Amend}, and purple for \textit{Monitor}. From left to right, the stages represent the full legislative process, from introduction to enactment, including committee review and the floor. The top chart shows the bill positions for the bills that are enacted, and the bottom chart shows the positions for the bills that are vetoed.

Overall, as the bill progresses, the frequency of lobbying tends to increase. However, the extent of lobbying by interest groups varies depending on their positions and the stage of the legislative process. This indicates that as the bill nears the enactment stage, lobbying efforts not only become more focused, with interest groups intensifying their activities to align the bill’s passage with their interests, but also the composition of dominant political groups shifts at each stage.

For the `Enacted' bills, the top line plot in Fig.~\ref{fig:bill_position_stage}, we observe a significant increase of \textit{Support} at the `Enacted' stage, while we observe minimal variation of \textit{Oppose} and \textit{Amend} across all stages.
We observe an increase of \textit{Monitor}, after the bill is enacted. After a bill becomes a law, it signals the implementation of its provisions, and as the bill often undergoes amendments afterward, many groups closely monitor the newly enacted law. This aligns with previous research~\citep{you2017ex}, which indicates that lobbying activities intensify post-enactment as interest groups aim to influence implementation and subsequent regulatory adjustments. Our study offers a direct characterization of this increase in lobbying, primarily driven by heightened \textit{Monitor} activities.

In contrast, for the `Vetoed' bills, the bottom line plot in Fig.~\ref{fig:bill_position_stage}, the patterns of \textit{Oppose} and \textit{Amend} differ significantly. The frequency of \textit{Oppose} and \textit{Amend} increases significantly after the bill passes the floor, and the bill ultimately fails.

Similarly, for the `Passed' and `Failed' bills during the floor process, we observe a comparable trend. In the case of Passed' bills, \textit{Monitor} activities increase towards the end, while in Failed' bills, \textit{Oppose} and \textit{Amend} activities rise. Detailed analysis is provided in the SI Appendix~\ref{sup:analysis1}.

Our research uncovers a systematic pattern in which active opposition to legislation emerges, particularly during the final stages of the legislative process. We suspect this may be due to a sense of urgency among stakeholders opposed to the bill, who view these final stages as their last chance to influence the outcome. Additionally, as a bill approaches enactment, it tends to attract more media attention and scrutiny. This increased visibility can mobilize opposition groups that may not have been previously aware of or concerned about the legislation.

\subsection*{Analysis 2: How Corporate Lobbying is related to Firm Size}
Firm size is widely recognized as a critical determinant of corporate political activities, primarily due to the fixed and variable costs associated with these efforts \citep{salamon1977economic, bombardini2008firm}. Although previous research indicates that only a few large firms engage in lobbying \citep{kim:17}, scholars have a limited understanding of whether firm size influences the decision to lobby and the nature of the lobbying positions adopted. To address this gap, we analyze lobbying decisions on two fronts: first, the decision to engage in lobbying, and second, the specific lobbying positions chosen by firms that do lobby.

We create a dataset that merges information on publicly traded companies from the Compustat database with lobbying records from 2009 to 2022. For each firm in each year, we first determine whether the firm engaged in lobbying. If it did, we compute the proportion of its lobbying activity for each position—\textit{Support}, \textit{Oppose}, \textit{Amend}, or \textit{Monitor}. To understand what influences a firm’s decision to lobby, we run logistic regression, using the number of employees as the primary measure of firm size. For firms that do lobby, we examine the relative composition of lobbying positions using Dirichlet regression, which is well-suited for analyzing proportional outcomes that sum to one.\footnote{The Dirichlet regression is estimated by maximizing the following log-likelihood function: $\ell(\beta) = \sum_{i=1}^{N} [\ln \Gamma(\sum_{j=1}^{J} \alpha_{ij}) - \sum_{j=1}^{J} \ln \Gamma(\alpha_{ij}) + \sum_{j=1}^{J} (\alpha_{ij} - 1) \ln(y_{ij})]$, where $\alpha_{ij} = \exp(x_i^\top \beta_j)$, $x_{ij}$ includes employment and year fixed effects, and $y_{ij}$ is the observed proportion of lobbying position $j$ for firm $i$.} For both models, our quantity of interest is defined as the average difference in predicted probabilities of lobbying outcomes between firms at the 90th and 10th percentiles of employment:  $\Delta = E[\hat{y}(emp^{90}) - \hat{y}(emp^{10})]$.

\begin{figure}[t]
    \centering
    \includegraphics[width=0.8\columnwidth]{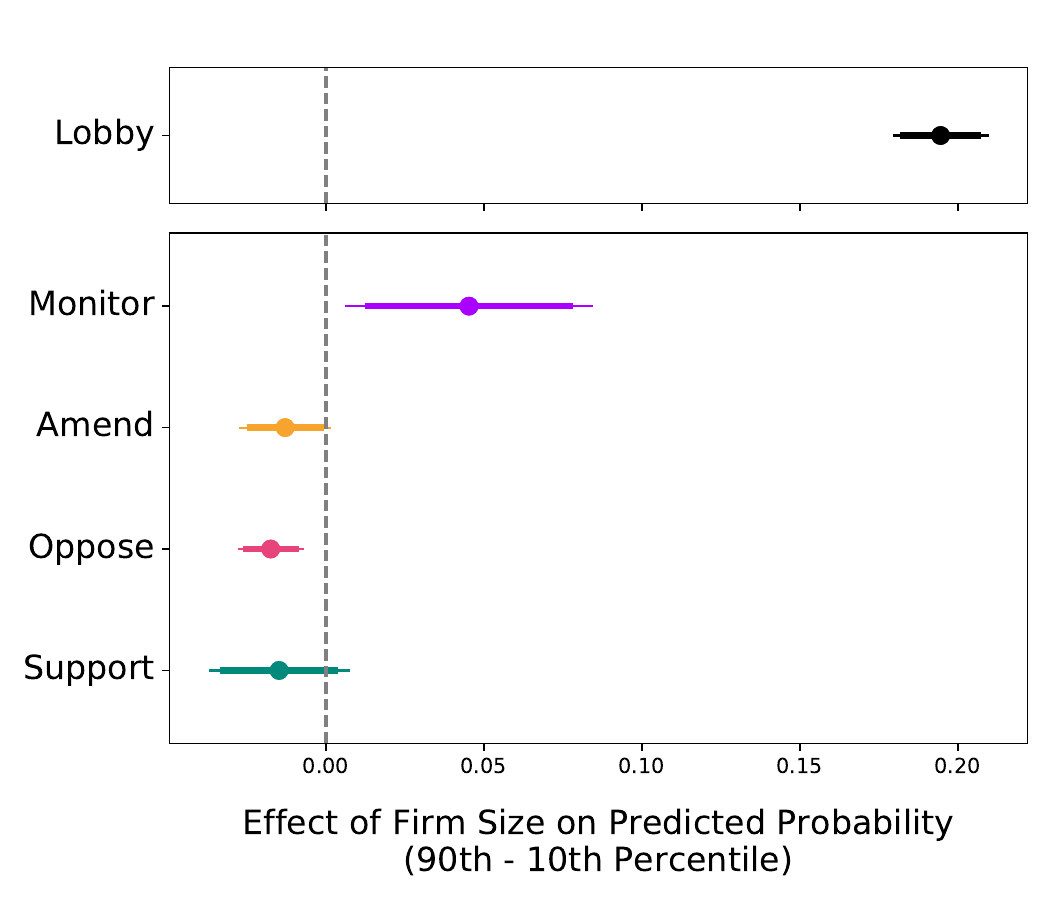}
    \caption{Predicted probability changes in lobbying pattern between firms at the 90th and 10th percentiles of employment. The top panel presents logistic regression–based estimates for the lobbying decision, while the bottom panel shows Dirichlet regression–based estimates for four lobbying positions (\textit{Support, Oppose, Amend, Monitor}). Each model uses firm‐year level data and includes year fixed effects, and error bars represent the 90\% and 95\% confidence intervals (thick and thin lines, respectively) obtained using block bootstrap at the firm level.}
    \label{fig:firm_size_prob}
\end{figure}

Fig.~\ref{fig:firm_size_prob} presents the regression results. The top panel shows that larger firms have a substantially higher propensity to engage in lobbying, consistent with the literature; firms at the 90th percentile of size are 19.5 percentage points (p.p.) more likely to lobby than those at the 10th percentile. The bottom panel reveals distinctive patterns in the specific lobbying positions adopted by firms of different sizes. In particular, larger firms are significantly more likely to adopt \textit{Monitor} positions (4.5 p.p difference), allowing them to track policy developments and build long-term relationships with policymakers. This strategic choice likely reflects their resource advantages, which enable sustained engagement in the policy process even without immediate policy objectives.

Conversely, smaller firms are more likely to adopt reactive stances, specifically through \textit{Oppose} or \textit{Amend} positions, though the magnitude of these changes is modest (differences of 1.8 and 1.3 p.p., respectively). This may reflect the resource disadvantages of smaller firms and aligns with our earlier findings that the frequency of opposition increases significantly at the final stage of legislation. On the other hand, we observe no statistical difference in the predicted probabilities for \textit{Support} positions. Notably, our findings---particularly the positive relationship for \textit{Monitor} positions---remain consistent across various model specifications that include additional controls such as industry fixed effects, fixed assets, and cost of goods sold, as detailed in the SI Appendix~\ref{sup:analysis2}.

Together, these results reveal how firm size shapes not only the decision to lobby but also the strategic positioning within the policy process, highlighting nuanced associations that call for further investigation into firm-level factors.

\begin{figure}[t]
\centering
\includegraphics[width=\linewidth]{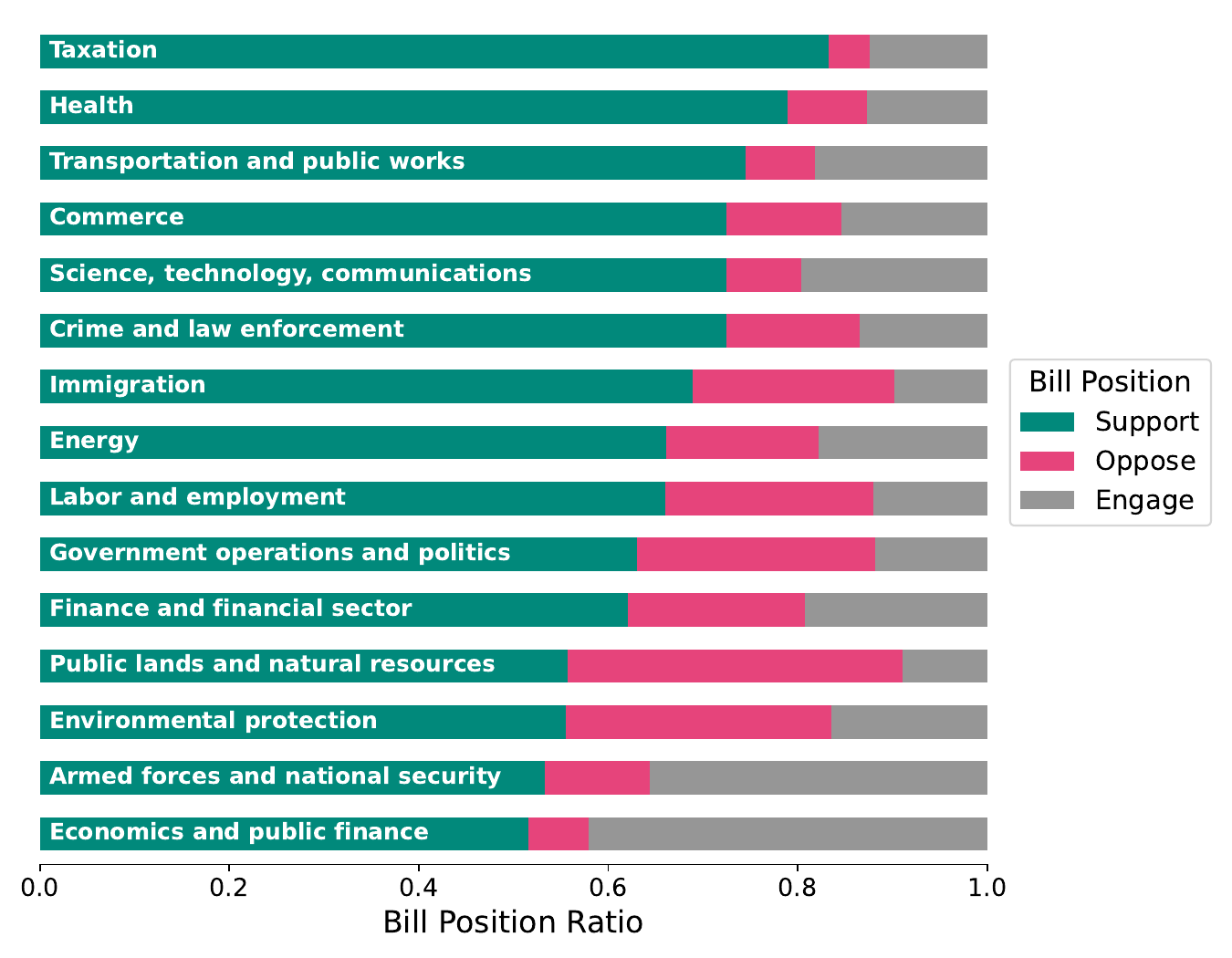}
\caption{Bill position ratios across different bill subjects. This shows the bill position ratios (\textit{Support}, \textit{Oppose}, and \textit{Engage}) for the top 15 bill subjects containing the largest number of bills. The subjects are ordered by their \textit{Support} ratio, illustrating how lobbying strategies vary across different legislative topics. \textit{Taxation}-related bills have the highest \textit{Support} ratio, indicating strong industry backing for favorable fiscal policies. In contrast, bills concerning \textit{Public lands and natural resources} exhibit a higher \textit{Oppose} ratio, reflecting resistance from industries affected by environmental regulations. Meanwhile, subjects such as \textit{Economics and public finance} show a greater proportion of \textit{Engage} positions, as lobbying in these cases often focuses on influencing appropriation decisions and securing favorable budget allocations rather than taking a definitive stance for or against the legislation.}
\label{fig:bill_position_ratio}
\end{figure}

\subsection*{Analysis 3: How Bill Positions Differ Across Subjects}
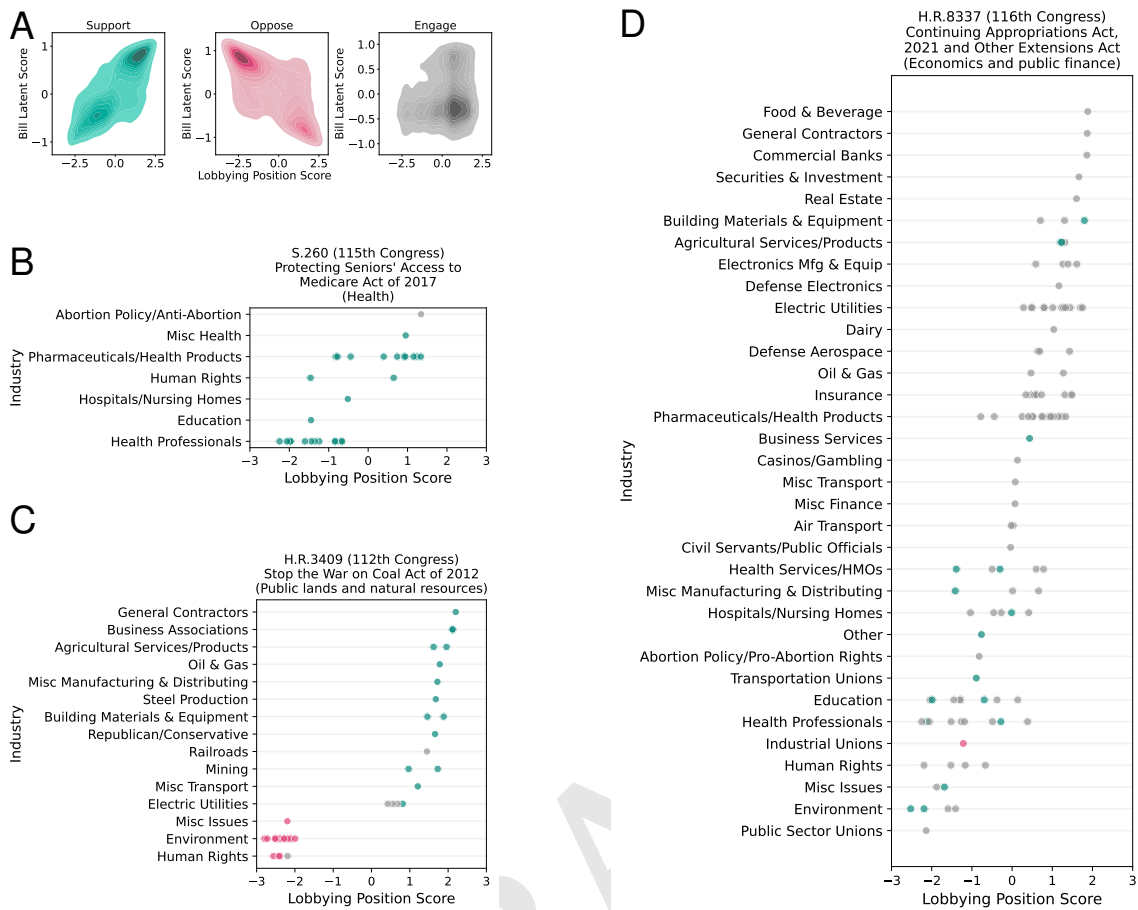
\begin{figure*}[!t]
    \centering
    \begin{minipage}[]{0.4\textwidth}
        \vskip 5mm
        \begin{tikzpicture}
            \node[inner sep=5pt] (image) {\includegraphics[width=0.9\textwidth]{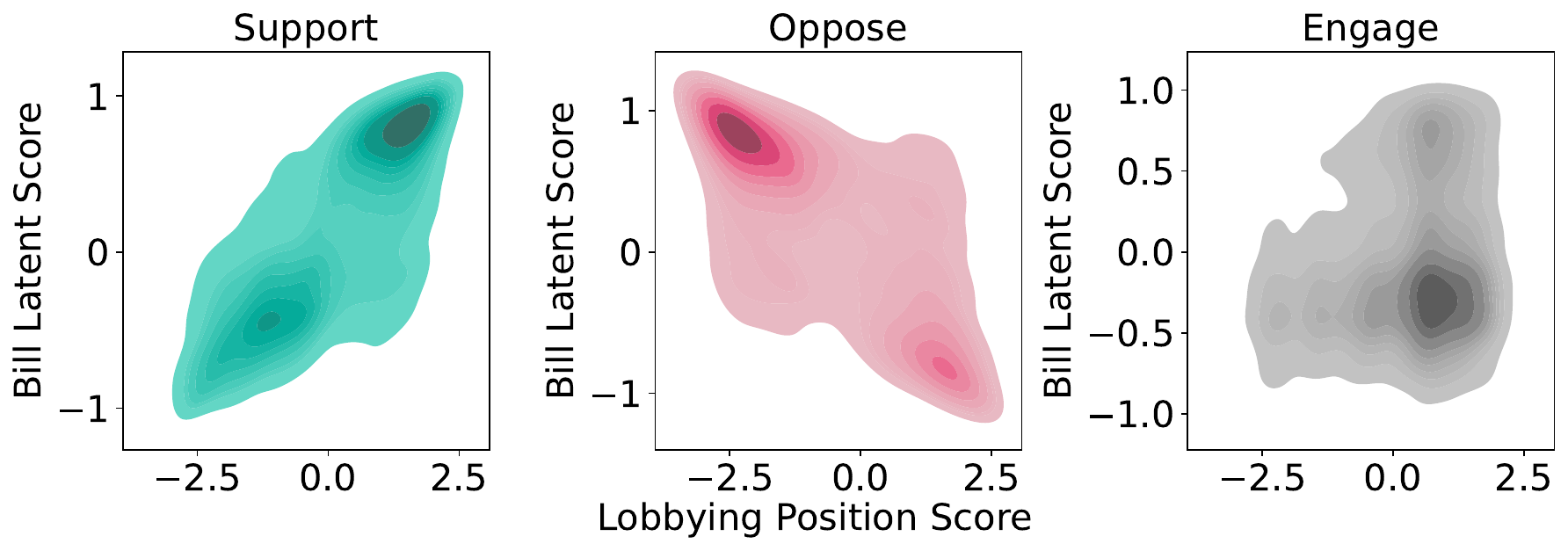}};
            \node[font={\fontsize{15}{5}\selectfont\sffamily}, anchor=north west] at (image.north west) {A};
        \end{tikzpicture}
        \vskip 5mm
        \begin{tikzpicture}
            \node[inner sep=0pt] (image) {\includegraphics[width=0.93\textwidth]{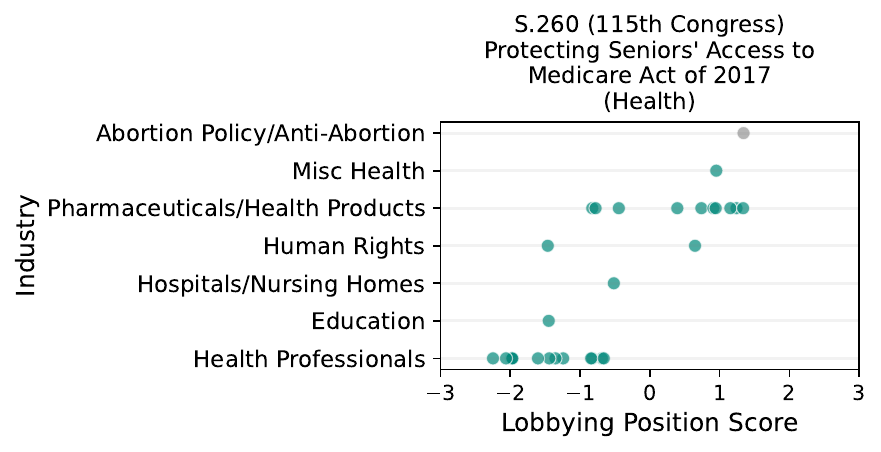}};
            \node[font={\fontsize{15}{5}\selectfont\sffamily}, anchor=north west] at (image.north west) {B};
        \end{tikzpicture}
        \begin{tikzpicture}
            \node[inner sep=0pt] (image) {\adjustbox{raise=5mm}{\includegraphics[width=0.93\textwidth]{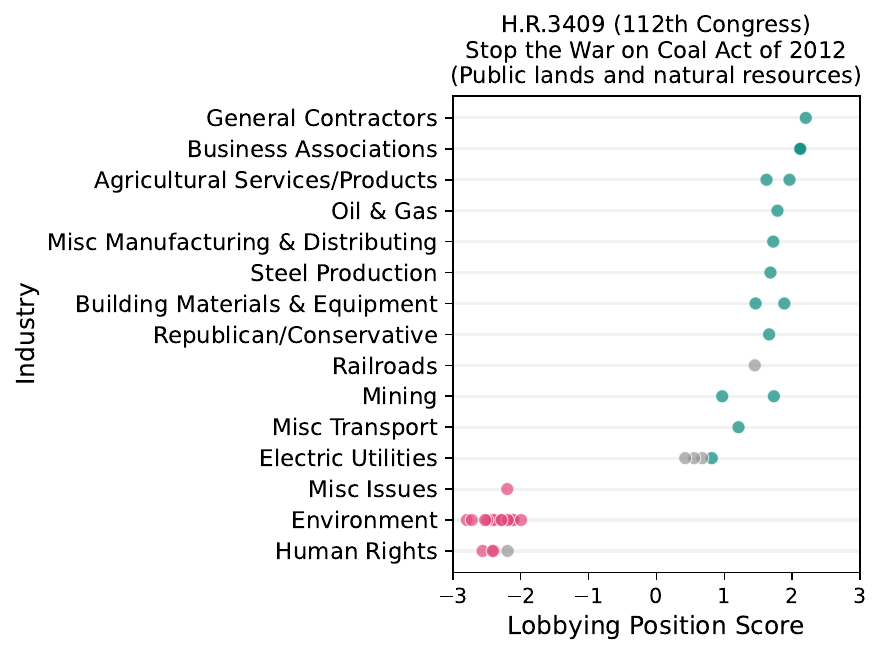}}};
            \node[font={\fontsize{15}{5}\selectfont\sffamily}, anchor=north west, yshift=6.5mm] at (image.north west) {C};
        \end{tikzpicture}
    \end{minipage}
    \begin{minipage}[]{0.5\textwidth}
        \centering
        \begin{tikzpicture}
            \node[inner sep=0pt] (image) {\includegraphics[width=0.8\textwidth]{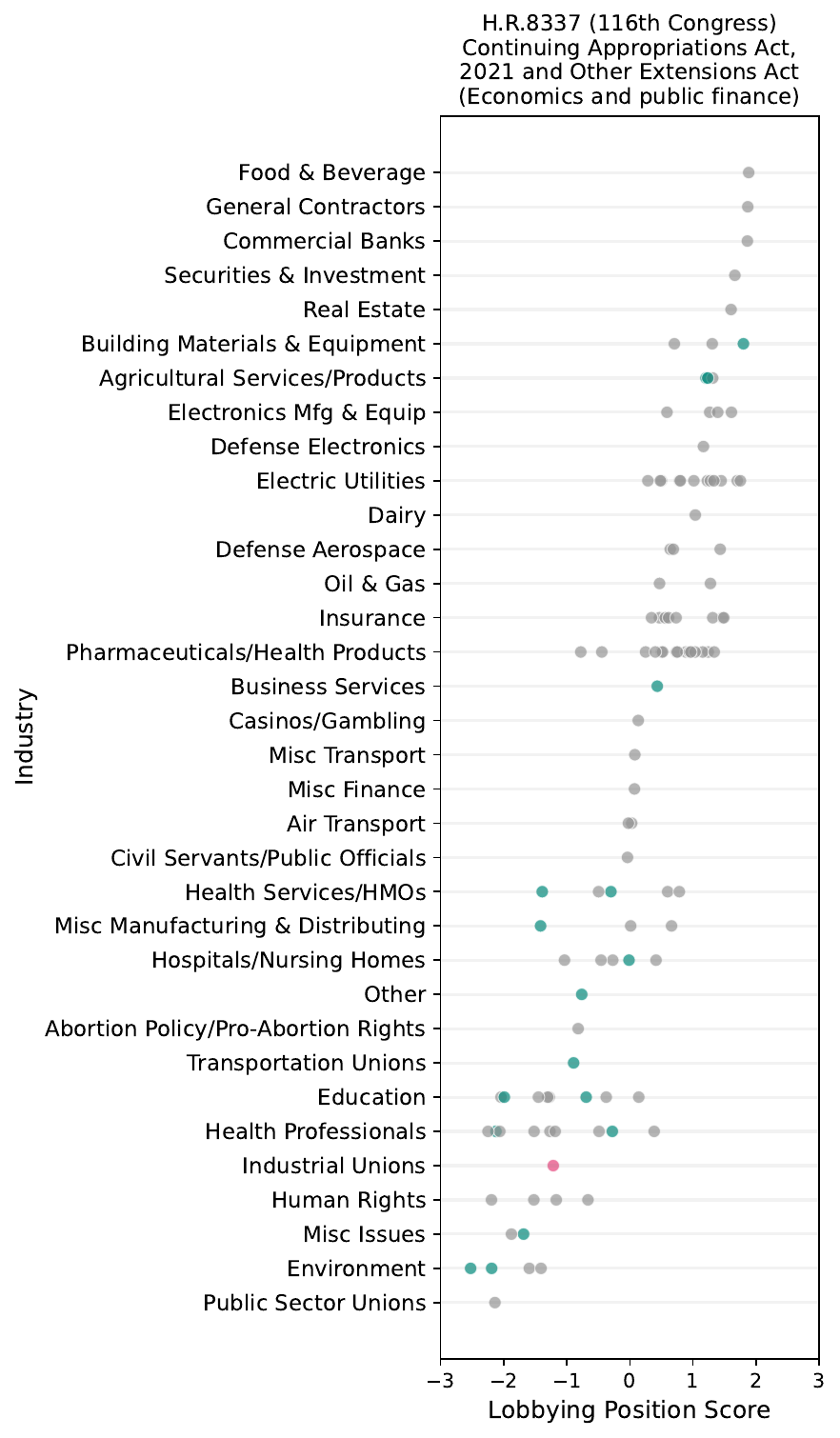}};
            \node[font={\fontsize{15}{5}\selectfont\sffamily}, anchor=north west] at (image.north west) {D};
        \end{tikzpicture}
    \end{minipage}
    \caption{ Bill-level distribution of bill positions across interest groups. (A) Distribution of bill positions across the latent scores of bills and interest groups. The x-axis represents the LPscore of interest groups, with commercial/conservative-leaning groups on the right and welfare/liberal-oriented groups on the left. The y-axis represents the latent scores of bills, where Republican-sponsored bills are generally positioned in the upper range and Democrat-sponsored bills in the lower range. \textit{Support} tends to occur when interest groups and bills have similar policy preferences, forming an upward diagonal pattern, while \textit{Oppose} is more frequent when they diverge, creating a downward diagonal trend. \textit{Engage (Amend/Monitor)} is more concentrated in the center without a distinct directional trend. This reveals distinct patterns in how bill positions vary based on the policy preferences of interest groups. 
    Panel (B, C, D): Comparison of bill position distribution by industry at the bill level. The y-axis represents the industries of interest groups, ordered by the descending average of all LP scores within each industry. We compare bill positions across different policy areas.
    (B) Taxation—Support is concentrated in the medical sector, with minimal opposition from other industries. (C) Public Lands and Natural Resources—A sharp divide—commercial and industry groups support the bill, while environmental and human rights groups strongly oppose it, reflecting conflicting economic and environmental interests.
    (D) Economics and Public Finance— The broad-impact economic bill attracts diverse interest groups, resulting in varied lobbying positions across industries. These patterns suggest that interest groups’ lobbying behavior and bill positions can be analyzed at the bill level, revealing sectoral alignment, industry conflicts, and the extent of interest group engagement across different policy areas.}
    \label{fig:bill_industry_bill_position}
\end{figure*}

Interest groups tend to adopt different positions on bills depending on the subject matter, with their strategies varying across policy areas. Fig.~\ref{fig:bill_position_ratio} shows the ratio of bill positions across the 15 most common bill subjects, ranked by \textit{Support} ratio. \textit{Support} is generally the most common position, while the proportions of \textit{Oppose} and \textit{Engage} vary by subject, reflecting differences in how interest groups engage with legislation. Several examples of topic-specific bills and associated interest group positions are illustrated in Fig.~\ref{fig:bill_industry_bill_position} (B), (C), and (D).

\paragraph{Bills with a high \textit{Support} ratio} 
Bill subjects with a high \textit{Support} ratio are often linked to specific policy areas, such as \textit{Taxation} and \textit{Health}. Major legislative initiatives in these categories typically involve tax reductions, financial incentives, or regulatory relief.
For example, several bills focus on specific population groups or goods (e.g., \texttt{S. 260 (115th) - Protecting Seniors' Access to Medicare Act of 2017} (Fig.~\ref{fig:bill_industry_bill_position} (B)); \texttt{S. 1562 (114th) - Reform Taxation of Alcoholic Beverages}). These bills benefit a clearly defined set of stakeholders, while they do not negatively affect anyone, thus it makes sense that there is minimal opposition and high support.
In some cases, we observe broader participation from multiple industries, particularly in legislation involving small business tax deductions (e.g., \texttt{S. 3612 (116th) - Small Business Expense Protection Act of 2020}), where various sectors benefit from government-backed financial relief.
Even when the policy targets a specific population, such as \texttt{H.R. 674 (112th) - Veterans' Pensions and Compensation}, the incentives offered can attract support from a diverse range of industries and interest groups.

\paragraph{Bills with a high \textit{Oppose} ratio} 
Conversely, legislative topics with relatively low \textit{Support} ratios are predominantly associated with public policy and regulatory matters. Notably, environmental and resource-related bills (e.g., \textit{Public lands and natural resources}, \textit{Environmental Protection}) exhibit a high \textit{Oppose} ratio, indicating interest group efforts to counteract regulatory changes. For instance, coal industry deregulation (e.g., \texttt{H.R. 3409 (112th) - Stop the War on Coal Act of 2012} (Fig.~\ref{fig:bill_industry_bill_position} (C)) and  \texttt{H.R. 806 (115th) - Ozone Standards Implementation Act of 2017}) are strongly opposed by environmental interest groups while supported by commercial entities. In contrast, environmental safety and protection legislation (e.g., \texttt{S. 847 (112th) - Safe Chemicals Act of 2011}) is supported by environmental interest groups, while opposed by commercial industries.

\paragraph{Bills with a high \textit{Engage} ratio} 
Bills with a high \textit{Engage} ratio mainly pertain to public-sector legislation, including \textit{Armed forces and national security} and \textit{Economics and public finance}. These topics frequently involve budget allocation, resulting in a broad set of stakeholders lobbying due to the diverse funding distribution across sectors. An example, \texttt{H.R. 8337 (116th) - Continuing Appropriations Act, 2021 and Other Extensions Act}) (Fig.~\ref{fig:bill_industry_bill_position} (D)) affects multiple stakeholders and attracts a larger number of lobbying entities per bill. 
Additionally, legislation concerning broad budget allocations (e.g., \texttt{H.R. 3082 (111th) - Continuing Appropriations and Surface Transportation Extensions Act, 2011}; \texttt{H.R. 133 (116th) - Consolidated Appropriations Act, 2021}) in defense, small businesses, transportation, rural development, energy, and environmental protection, attract lobbying by a diverse set of interest groups.

Further details on the bills and the bill positions of the interest groups can be found in the SI Appendix~\ref{sup:analysis3}.

\begin{figure}[t]
\centering
\includegraphics[width=\linewidth]{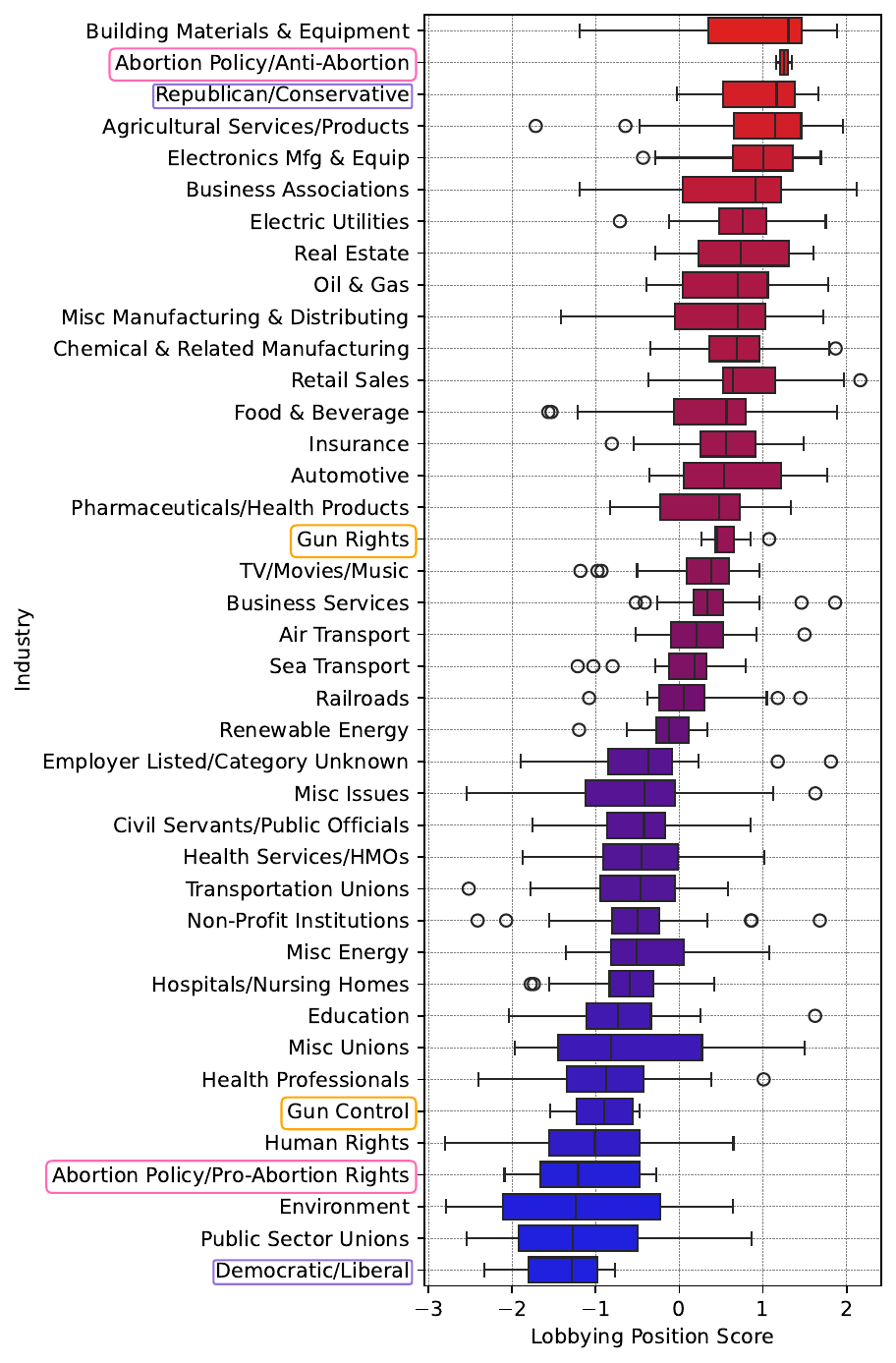}
\caption{Industry-wise distribution of interest group preferences. The boxplot illustrates the distribution of LPscores for 1,414 interest groups across 40 industries. Industries are sorted by average scores in descending order, with economic and conservative-aligned sectors at the top and social/progressive sectors at the bottom. Conflicting interest industries (e.g., \textit{Republican} vs. \textit{Democratic}, \textit{Pro-Abortion} vs. \textit{Anti-Abortion}, \textit{Gun Rights} vs. \textit{Gun Control}) are distinctly positioned, highlighting ideological divides.}
\label{fig:boxplot_industry}
\end{figure}

\subsection*{Analysis 4: How Interest Group Preferences Vary Across Industries}
The interest group's latent score reflects their hidden preferences. Since each interest group has its own distinct interests, we compare the distribution of LPscores across industries to examine the similarities and differences in their preferences.
Fig.~\ref{fig:boxplot_industry} presents the LPscore distribution of 1,414 interest groups, covering 40 industries among the total 94. This includes the 36 industries,\footnote{Industries are grouped according to the classification defined by OpenSecrets. For more details, please refer to \url{https://www.opensecrets.org/federal-lobbying/industries}}  with the most common interest groups out of 94, adding three industry pairs—Abortion, Gun, and Ideology—that exhibit conflicting interests, visually grouped using the same box color.
Industries in Fig.~\ref{fig:boxplot_industry} are ordered by the average LP score in descending order from top to bottom.
Overall, industries at the top are primarily associated with economic and industrial activities or aligned with conservative ideologies, such as \textit{Building Materials \& Equipment}, \textit{Agricultural Services/Products}, \textit{Real Estate}, and \textit{Oil \& Gas}. On the other hand, those at the bottom are more focused on social values and services or tied to progressive ideologies, including sectors like \textit{Public Sector Unions}, \textit{Environment}, \textit{Human Rights}, and \textit{Education}.
By comparing the positions of the 6 industries with conflicting interests, \textit{Republican/Conservative}, \textit{Abortion Policy/Anti-Abortion}, and \textit{Gun Rights} industries are placed at the top, while \textit{Democratic/Liberal}, \textit{Abortion Policy/Pro-Abortion}, and \textit{Gun Control} industries are positioned at the bottom.  This clear separation highlights the latent divide in interests, effectively capturing the contrasting perspectives of these opposing groups.

\paragraph{How does the distribution of interest groups' bill positions vary based on the ideology of the bill?}
Fig.~\ref{fig:bill_industry_bill_position} (A) illustrates the kernel density estimation (KDE) plots showing the distribution density of bill positions based on the bills and interest groups.\footnote{The KDE plots are generated using the kdeplot function from the Seaborn library. See \url{https://seaborn.pydata.org/generated/seaborn.kdeplot.html}}
The x-axis represents LPscore of interest groups, while the y-axis corresponds to the latent scores of bills. Bills sponsored by Republicans tend to be positioned higher on the y-axis, while those sponsored by Democrats are generally lower. (Further details can be found in the SI Appendix~\ref{sup:analysis4}.) On the x-axis, interest groups with business or conservative leanings are more prevalent toward the right, whereas welfare- and liberal-oriented groups are more common on the left.
Bill positions vary depending on both the interest group and the bill. 
\textit{Support} tends to appear where interest groups and bills have similar policy preferences, forming an upward diagonal pattern. \textit{Oppose} is more common where their preferences diverge, creating a downward diagonal trend. \textit{Engage}, including \textit{Amend} and \textit{Monitor}, are more centrally distributed without a strong directional tendency. This shows how bill positions reflect interest groups' policy alignment or divergence, providing insight into their interests.

\paragraph{How do interest groups differ in their bill positions}
Our data allow us to analyze interest groups' bill positions at the bill level. Fig.~\ref{fig:bill_industry_bill_position} (B), (C), and (D) illustrate examples of interest groups' bill positions on specific bills. The dots along the lines represent interest groups within each industry.

In Fig.~\ref{fig:bill_industry_bill_position} (B), we examine \texttt{S. 260 (115th) - Protecting Seniors’ Access to Medicare Act of 2017}, a bill related to \textit{Taxation}. Consistent with the taxation bill ratio in Fig.~\ref{fig:bill_position_ratio}, only a small subset of interest groups—mainly those associated with the medical sector—engaged in lobbying on this bill, with a strong tendency to support it.

Fig.~\ref{fig:bill_industry_bill_position} (C) presents \texttt{H.R.3409 (112th) - Stop the War on Coal Act of 2012}, which falls under the \textit{Public Lands and Natural Resources} category. This bill has a high proportion of opposition, reflecting a clear conflict of interest among industries. Interest groups positioned at the top of the graph belong to the \textit{Commercial \& Industry Sectors}, while those at the bottom are affiliated with \textit{Environmental} and \textit{Human Rights} groups. The stark division in lobbying positions highlights the competing economic and environmental interests surrounding this legislation.

Finally, Fig.~\ref{fig:bill_industry_bill_position} (D) showcases \texttt{H.R.8337 (116th) - Continuing Appropriations Act, 2021 and Other Extensions Act}, which belongs to \textit{Economics and Public Finance}. Bills of this nature typically attract interest groups from a wide array of industries, as they encompass diverse topics with broad economic and regulatory implications. Consequently, these bills tend to have a higher proportion of \textit{Engage} as their lobbying position.

These examples illustrate how interest groups' bill positions are shaped by industry-specific interests, with some bills attracting support from niche sectors, others revealing stark conflicts between opposing industries, and broad-reaching legislation engaging a wide array of stakeholders.

These examples illustrate how bill positions vary across interest groups. While some bills see strong alignment within specific sectors, others expose divisions between industries, reflecting the complexity of policy preferences in lobbying 
dynamics.

\section*{Conclusion}
This study presents a novel, large-scale approach to measure and analyze the positions of special interest groups (SIGs) in the U.S. legislative process. By leveraging advanced AI methods—Large Language Models (LLMs) and Graph Neural Networks (GNNs)—we developed an automated and scalable pipeline that extracts bill positions from lobbying activities. This framework enables a more systematic and fine-grained understanding of how interest groups shape legislation beyond traditional datasets that rely on indirect measures such as campaign finance data.

Our findings highlight several key insights into the lobbying landscape. First, there is a strong correlation between a bill’s legislative progression and the positions taken by interest groups, illustrating their strategic engagement at different stages. Second, firm size plays a critical role in lobbying decisions, with larger firms tending to adopt monitoring positions,  leveraging their resource advantages, while smaller firms show a relatively greater reliance on reactive oppose and amend positions. Third, the distribution of bill positions varies significantly across legislative topics, revealing that industries align their lobbying efforts based on issue-specific stakes. Finally, interest groups’ lobbying behavior and bill positions can be analyzed at the bill level, revealing sectoral alignments, industry conflicts, and the extent of interest group engagement across different policy areas, with variations observed across both interest groups and specific policy domains.

By providing a more comprehensive and structured dataset, this research paves the way for deeper investigations into how interest groups influence legislative outcomes, the interplay between corporate interests and public policy, and the broader implications of lobbying on democratic governance. Moreover, our AI-driven approach offers a scalable model for future studies, allowing researchers and policymakers to track lobbying activities more effectively over time. As artificial intelligence continues to enhance political science research, our work underscores its potential in increasing transparency and understanding in legislative processes. Future research can expand upon this foundation by incorporating additional data sources, refining predictive models, and exploring the evolving strategies of interest groups in shaping public policy.








\acknow{
This work was supported by MIT International Science and Technology Initiatives (MISTI) Global Seed Fund. }

\showacknow{} 


\bibliography{main-bib}

\begin{thebibliography}{10}

\bibitem{baumgartner2009lobbying}
FR Baumgartner, JM Berry, M Hojnacki, BL Leech, DC Kimball, {\em Lobbying and policy change: Who wins, who loses, and why}.
\newblock (University of Chicago Press), (2009).

\bibitem{grossman:helpman:01}
GM Grossman, E Helpman, {\em Special interest politics}.
\newblock (MIT Press, Cambridge, MA, USA), (2001).

\bibitem{hall:dear:06}
RL Hall, AV Deardorff, Lobbying as legislative subsidy.
\newblock {\em\protect\JournalTitle{American Political Science Review}} \textbf{100}, 69--84 (2006).

\bibitem{kim:17}
IS Kim, Political cleavages within industry: Firm level lobbying for trade liberalization.
\newblock {\em\protect\JournalTitle{American Political Science Review}} \textbf{111}, 1--20 (2016).

\bibitem{farrell2016pnas}
J Farrell, Corporate funding and ideological polarization about climate change.
\newblock {\em\protect\JournalTitle{Proceedings of the National Academy of Sciences}} \textbf{113}, 92--97 (2016).

\bibitem{brulle2014institutionalizing}
RJ Brulle, Institutionalizing delay: foundation funding and the creation of us climate change counter-movement organizations.
\newblock {\em\protect\JournalTitle{Climatic change}} \textbf{122}, 681--694 (2014).

\bibitem{stokes2020short}
LC Stokes, {\em Short circuiting policy: Interest groups and the battle over clean energy and climate policy in the American States}.
\newblock (Oxford University Press), (2020).

\bibitem{xie2025tracing}
JJ Xie, et~al., Tracing inclusivity at unfccc conferences through side events and interest group dynamics.
\newblock {\em\protect\JournalTitle{Nature Climate Change}} pp. 1--9 (2025).

\bibitem{melissa2017pnas}
ML Sands, Exposure to inequality affects support for redistribution.
\newblock {\em\protect\JournalTitle{Proceedings of the National Academy of Sciences}} \textbf{114}, 663--668 (2017).

\bibitem{dietze2021framing}
P Dietze, MA Craig, Framing economic inequality and policy as group disadvantages (versus group advantages) spurs support for action.
\newblock {\em\protect\JournalTitle{Nature Human Behaviour}} \textbf{5}, 349--360 (2021).

\bibitem{wang2024global}
D Wang, Y Fang, Global climate governance inequality unveiled through dynamic influence assessment.
\newblock {\em\protect\JournalTitle{npj Climate Action}} \textbf{3}, 75 (2024).

\bibitem{milner:ting:15}
HV Milner, D Tingley, {\em Sailing the Water's Edge: The Domestic Politics of American Foreign Policy}.
\newblock (Princeton University Press, Princeton, NJ, USA), (2015).

\bibitem{liao:2023}
S Liao, The effect of firm lobbying on high-skilled visa adjudication.
\newblock {\em\protect\JournalTitle{The Journal of Politics}} \textbf{85}, 1416--1429 (2023).

\bibitem{kim2021mapping}
IS Kim, D Kunisky, Mapping political communities: A statistical analysis of lobbying networks in legislative politics.
\newblock {\em\protect\JournalTitle{Political Analysis}} \textbf{29}, 317--336 (2021).

\bibitem{grasse2011influence}
N Grasse, B Heidbreder, The influence of lobbying activityin state legislatures: Evidence from wisconsin.
\newblock {\em\protect\JournalTitle{Legislative Studies Quarterly}} \textbf{36}, 567--589 (2011).

\bibitem{daniel2022prq}
DM Butler, DR Miller, Does lobbying affect bill advancement? evidence from three state legislatures.
\newblock {\em\protect\JournalTitle{Political Research Quarterly}} \textbf{75}, 547--561 (2022).

\bibitem{bonica2014mapping}
A Bonica, Mapping the ideological marketplace.
\newblock {\em\protect\JournalTitle{American Journal of Political Science}} \textbf{58}, 367--386 (2014).

\bibitem{bonica2019donation}
A Bonica, Are donation-based measures of ideology valid predictors of individual-level policy preferences?
\newblock {\em\protect\JournalTitle{The Journal of Politics}} \textbf{81}, 327--333 (2019).

\bibitem{lorenz2020large}
GM Lorenz, AC Furnas, JM Crosson, Large-n bill positions data from maplight. org: What can we learn from interest groups’ publicly observable legislative positions?
\newblock {\em\protect\JournalTitle{Interest Groups \& Advocacy}} \textbf{9}, 342--360 (2020).

\bibitem{butl:mill:22}
DM Butler, DR Miller, Does lobbying affect bill advancement? evidence from three state legislatures.
\newblock {\em\protect\JournalTitle{Political Research Quarterly}} \textbf{75}, 547--561 (2022).

\bibitem{abi2023ideologies}
S Abi-Hassan, JM Box-Steffensmeier, DP Christenson, AR Kaufman, B Libgober, The ideologies of organized interests and amicus curiae briefs: Large-scale, social network imputation of ideal points.
\newblock {\em\protect\JournalTitle{Political Analysis}} \textbf{31}, 396--413 (2023).

\bibitem{poole1985spatial}
KT Poole, H Rosenthal, A spatial model for legislative roll call analysis.
\newblock {\em\protect\JournalTitle{American journal of political science}} pp. 357--384 (1985).

\bibitem{clinton2004statistical}
J Clinton, S Jackman, D Rivers, The statistical analysis of roll call data.
\newblock {\em\protect\JournalTitle{American Political Science Review}} \textbf{98}, 355--370 (2004).

\bibitem{samejima1968estimation}
F Samejima, Estimation of latent ability using a response pattern of graded scores 1.
\newblock {\em\protect\JournalTitle{ETS Research Bulletin Series}} \textbf{1968}, i--169 (1968).

\bibitem{chalmers2012mirt}
RP Chalmers, mirt: A multidimensional item response theory package for the r environment.
\newblock {\em\protect\JournalTitle{Journal of statistical Software}} \textbf{48}, 1--29 (2012).

\bibitem{bonica2023database}
A Bonica, Database on ideology, money in politics, and elections: Public version 3.1 [computer file].
\newblock {\em\protect\JournalTitle{URL: https://data. stanford. edu/dime}} (2023).

\bibitem{Holyoke02102019}
TT Holyoke, Strategic lobbying to support or oppose legislation in the us congress.
\newblock {\em\protect\JournalTitle{The Journal of Legislative Studies}} \textbf{25}, 533--552 (2019).

\bibitem{holyoke2022strategic}
TT Holyoke, Strategic lobbying and the pressure to compromise member interests.
\newblock {\em\protect\JournalTitle{Political Research Quarterly}} \textbf{75}, 1255--1270 (2022).

\bibitem{you2017ex}
HY You, Ex post lobbying.
\newblock {\em\protect\JournalTitle{The Journal of Politics}} \textbf{79}, 1162--1176 (2017).

\bibitem{salamon1977economic}
LM Salamon, JJ Siegfried, Economic power and political influence: The impact of industry structure on public policy.
\newblock {\em\protect\JournalTitle{American political science review}} \textbf{71}, 1026--1043 (1977).

\bibitem{bombardini2008firm}
M Bombardini, Firm heterogeneity and lobby participation.
\newblock {\em\protect\JournalTitle{Journal of International Economics}} \textbf{75}, 329--348 (2008).

\bibitem{wilkerson2015tracing}
Tracing the flow of policy ideas in legislatures: A text reuse approach.
\newblock {\em\protect\JournalTitle{American Journal of Political Science}} \textbf{59}, 943--956 (2015).

\bibitem{linder2020text}
F Linder, B Desmarais, M Burgess, E Giraudy, Text as policy: Measuring policy similarity through bill text reuse.
\newblock {\em\protect\JournalTitle{Policy Studies Journal}} \textbf{48}, 546--574 (2020).

\bibitem{kim-etal-2021-learning}
J Kim, E Griggs, IS Kim, A Oh, Learning bill similarity with annotated and augmented corpora of bills in {\em Proceedings of the 2021 Conference on Empirical Methods in Natural Language Processing}, eds.{} MF Moens, X Huang, L Specia, SWt Yih.
\newblock (Association for Computational Linguistics, Online and Punta Cana, Dominican Republic), pp. 10048--10064 (2021).

\end{thebibliography}

\section*{Supporting Information Appendix (SI)}
\section{LLM Annotation}
\label{sup:llm_annotation}

\begin{table*}[b]
    \centering
    \caption{Bill information and lobbying description for LLM annotation}
    \label{tab:llm_input_example}
    \begin{tabular}{>{\centering}p{0.7cm} p{2cm} p{4cm} p{5cm} c}
        Bill ID  &  Bill Short Title & Bill Official Title & Lobbying Description & Bill Position \\ 
        \midrule
        H.R.1421 &  Improving Access to Medicare Coverage Act of 2017  
                & To amend title XVIII of the Social Security Act to count a period of receipt of outpatient observation services in a hospital toward satisfying the 3-day inpatient hospital stay requirement for coverage of skilled nursing facility services under Medicare, and for other purposes. 
                & S. 568, H.R. 1421, Improving Access to Medicare Coverage Act - Discussed our support for/endorsed bill. & Support  \\ \cmidrule(lr){1-5} 

        H.R.1249 &  Leahy-Smith America Invents Act  
                & To amend title 35, United States Code, to provide for patent reform.  
                & defeat S. 23 and H.R. 1249, and pass funding only bill for USPTO. & Oppose  \\ \cmidrule(lr){1-5} 

        H.R.9    &  Innovation Act  
                & To amend title 35, United States Code, and the Leahy-Smith America Invents Act to make improvements and technical corrections, and for other purposes.  
                & Advocacy for changes to H.R. 9, the Innovation Act regarding the potential impact of provisions on the ability of universities to protect the rights of faculty and support research spin-outs. & Amend  \\ \cmidrule(lr){1-5} 

        S.744    &  Border Security, Economic Opportunity, and Immigration Modernization Act  
                & A bill to provide for comprehensive immigration reform and for other purposes.  
                & Educated members of Congress and administration officials about the need to reform the immigration process and expand the number of worker-sponsored visas for highly skilled workers. Legislation watched includes: S 169, S 744, HR 2131, HR 459, HR 15 and S 600. & Monitor  \\ 
        \bottomrule
    \end{tabular}
\end{table*}
\subsection*{Keyword extraction}
Table~\ref{tab:keyword_list} presents a comprehensive list of keywords related to bill positions, which are essential for identifying and categorizing lobbying activities linked to specific legislative bills. These keywords are carefully selected based on established definitions of lobbying as outlined in state statutes and guidance documents,\footnote{These definitions include terms related to lobbying and lobbyists, as outlined in the statutes of the respective states. See \url{www.sos.state.co.us/pubs/lobby/files/guidanceManual.pdf}} including the \textit{Colorado Secretary of State Lobbying Guidance Manual}.\footnote{For reference, we use the \textit{Colorado Secretary of State Lobbying Guidance Manual}. See \url{www.ncsl.org/ethics/how-states-define-lobbying-and-lobbyist}} 
They represent terms that denote various positions or actions that interest groups take in relation to legislation, including advocating for, supporting, opposing, or monitoring bills. 
The keywords in this list—such as \textit{support}, \textit{endorse}, \textit{oppose}, \textit{defeat}, and \textit{watch}—are specifically chosen because they reflect the primary lobbying stances and activities that are tracked and reported by interest groups.
By focusing on these key terms, we efficiently filter and segment the large volume of lobbying data, ensuring that only relevant bill-specific lobbying activities are included for further analysis. 
This structured approach allows for a more targeted and effective analysis of how interest groups engage with specific bills, ultimately enhancing the accuracy and usefulness of the dataset.

\begin{table}[H]
\centering
\caption{Keyword list related to bill positions}
\label{tab:keyword_list}
\resizebox{0.45\textwidth}{!}{
\begin{tabular}{p{0.45\textwidth}}
\multicolumn{1}{c}{Bill Position Keywords} \\ \midrule
support, advocate, favor, endorse, engage, encourage, yes, passage, promote, attempt, influence, introduce, propose, draft, oppose, against, defeat, reject, repeal, amend, modify, delete, monitor, track, watch \\
\bottomrule
\end{tabular}
}
\end{table}

\subsection*{LLM Configuration}
We use the GPT-4 model for LLM annotation, utilized through Azure OpenAI. The specific specifications of the model are provided in Table~\ref{tab:llm_setting}.
The temperature is set to 0 to ensure deterministic responses.
\begin{table}[H]
\centering
\caption{GPT-4 setting}
\label{tab:llm_setting}
\begin{tabular}{cc}
Setting     & Value              \\ \midrule
Temperature & 0.0                \\
Model       & gpt-4-1106-Preview \\
API version & 2023-09-01-preview \\ \bottomrule
\end{tabular}
\end{table}

\begin{table*} [t]
\centering
\caption{Annotation Guideline}
\label{tab:annotation_guide}
\begin{tabular}{cp{13cm}}
Bill Position & \multicolumn{1}{c}{Description}                                                                                                            \\ \midrule
Support       & If the lobbying activity explicitly supports the bill, label it as 'Support'. Efforts to pass the bill through lobbying, endorsing specific parts of the bill, proposing the bill, or participating in its introduction all fall under the category of 'Support'. \\ \cmidrule(lr){1-2} 
Oppose        & If the lobbying activity explicitly supports the bill, label it as 'Oppose'. If the activity opposes specific parts of the bill, it should also be labeled as 'Oppose'. \\ \cmidrule(lr){1-2} 
Amend         & For cases labeled 'Amend': Label as 'Amend' if the lobbying activity aims to change the bill. 'Amend' is only applicable when there is explicit mention of intending to amend the bill. Additionally, mentioning an amendment to the bill is also considered as 'Amend'. For cases labeled ' Mention': Label as 'Mention' if the lobbying activity does not aim to change the bill. If the bill itself is aimed at amending another bill, and the description simply outlines the content of the bill, this does not constitute amending the bill, so it should be annotated as 'Mention'. \\ \cmidrule(lr){1-2} 
Monitor       & If the lobbying activity is explicitly stated to involve monitoring or tracking the bill, label it as 'Monitor'. Taking a position (supporting, opposing) or engaging in lobbying for amendments to the bill does not classify as 'Monitor'. \\ \bottomrule                                                                                                        
\end{tabular}
\end{table*}

\subsection*{Bill Position LLM Prompt}
To automatically annotate large-scale bill positions, lobbying activity texts are classified using large language models (LLMs). The LLM classifies the descriptions into one of five categories: \textit{Support}, \textit{Oppose}, \textit{Amend}, \textit{Monitor}, or \textit{Mention}. We use the following prompt to classify lobbying activity texts:
\begin{center}
\fbox{
  \centering
  \begin{minipage}{0.9\columnwidth}
    {Classify the given text that explicitly describes lobbying activities for bill \{\textbf{bill\_id}\} into one of the five types: \textit{Support}, \textit{Oppose}, \textit{Amend}, \textit{Monitor}, or \textit{Mention} without explanation.\\\\
    \{\textbf{bill\_id}\} short title: \{\textbf{short\_title}\}\\
    \{\textbf{bill\_id}\} official title: \{\textbf{official\_title}\}\\
    Text: \{\textbf{lobbying\_description}\}\\
    Answer:}
  \end{minipage}
}
\end{center}
The prompt includes the five categories and provides both the short and official titles of the bill to offer context about its subject. It also includes the lobbying description associated with the bill. To ensure that only the class label is outputted, without additional explanations, we append the instruction ``without explanation'', as LLMs usually provide reasoning along with their answers. We try multiple versions with more detailed information, but the simplest version consistently yields high performance, so we choose this version.

\subsection*{LLM Input Example}
Table~\ref{tab:llm_input_example} below provides an example of the bill ID, bill short title, bill official title, and lobbying description that are input into the LLM prompt. The result shows how the LLM accurately outputs the corresponding bill position based on this input.


\subsection*{Human Annotation Guidelines for Bill Position Classification} To validate the LLM annotation, we established human annotation guidelines as shown in Table~\ref{tab:annotation_guide}. The two authors, who thoroughly understand the bill position classification, followed these guidelines to annotate validation samples. In this guideline, when a lobbying description from a lobbying report clearly states one of the four bill positions (\textit{Support}, \textit{Oppose}, \textit{Amend}, \textit{Monitor}), it is classified under the corresponding category.

\section{GNN Annotation}
\label{sup:gnn_annotation}

\begin{figure*}[t]
\centering
\includegraphics[width=0.85\textwidth]{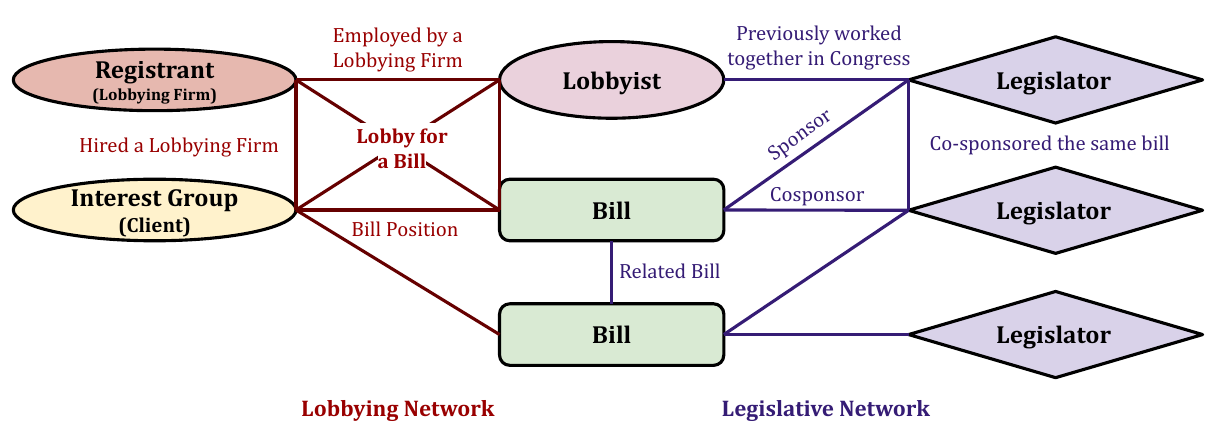}
\caption{Graph Structure Overview of Lobbying and Legislative Networks. The lobbying network includes interest groups, registrants (such as lobbying firms), and lobbyists. The legislative network captures relationships among bills, sponsors, cosponsors, and lobbyists who have worked with specific legislators. These networks are used to infer interest group bill positions by leveraging their connections and interactions.}
\label{fig:graph_structure}
\end{figure*}

\subsection*{Graph Structure Overview: Lobbying and Legislative Networks}
When the textual content of lobbying activity alone is not sufficient to fully capture the nuances of interest group bill positions, we enhance the classification by leveraging both the lobbying and legislative networks.
As illustrated in Fig.~\ref{fig:graph_structure}, these networks consist of multiple types of entities. The lobbying network includes interest groups, registrants (e.g., lobbying firms) who register lobbying activities and reports on behalf of those interest groups, and lobbyists affiliated with those firms who lobby on specific bills.
The legislative network captures relationships among bills, legislators who sponsor or cosponsor them, and includes additional connections between lobbyists and legislators when a lobbyist has previously worked with a specific member of the U.S. Congress.
Additionally, we involve \textit{Related Bill} edges in the graph to represent complex inter-bill relationships, such as overlapping content,\footnote{Related bill information from \url{www.congress.gov}. See \url{https://www.congress.gov/help/related-bills}} which provide important context for understanding the legislative process~\citep{wilkerson2015tracing, linder2020text, kim-etal-2021-learning}.

\subsection*{Entity Features}
Our graph is a heterogeneous structure that includes multiple entity types. Each entity type has different features, which are embedded as nodes. Table~\ref{tab:graph_node_feature} provides the number of entities of each entity type in the overall graph, the total embedding size, the features included, the embedding size of each feature, and examples of the features. For features such as \textit{Industry}, \textit{State \& Territories}, and \textit{Government}, which are represented as binary or multi-class, embeddings are created using one-hot encoding. 
For text, features include company/lobbyist names, their brief descriptions, bill titles, and bill summaries. These texts are first stemmed (a process of reducing words to their root forms) using the Porter Stemmer,\footnote{See~\url{https://www.nltk.org/api/nltk.stem.porter.html\#nltk.stem.porter.PorterStemmer}} and then tf-idf (Term Frequency-Inverse Document Frequency), which captures the importance of key terms in the context of a document, is used to create a 64-dimensional embedding with a word dictionary, implemented via the TfidfVectorizer package.\footnote{See~\url{https://scikit-learn.org/stable/modules/generated/sklearn.feature_extraction.text.TfidfVectorizer.html}} For more detailed information, please refer to our data.

\begin{table*}[t]
\centering
\caption{Node Features in Graph}
\label{tab:graph_node_feature}
\resizebox{0.9\textwidth}{!}{
\begin{tabular}{crrlrl}
Entity Node    & The Number of Entity & Emedding Size & Feature Name             & Feature Size & Example                                                       \\ \midrule
Interest Group & 16093                & 557           & Industry                 & 426          & Agriculture, Crop production \& basic processing,...              \\
               &                      &               & State\&Territories       & 59           & NY, AL, ... + U.S. territories (DC, PR, GU, ...)              \\
               &                      &               & Government               & 2            & True, False                                                   \\
               &                      &               & Industry Type            & 6            & org, com, edu, net, gov, others                               \\
               &                      &               & Text (Name, Description) & 64           & advocaci, allianc, america, health, manufactur, ..            \\ \cmidrule(lr){1-6}
Bill           & 57466                & 114           & Subject                  & 34           & Crime and law enforcement,,...       \\
               &                      &               & Final State              & 9            & ENACTED:SIGNED, FAIL, PASSED:BILL,...               \\
               &                      &               & Party                    & 3            & Democrat, Independent, Republican                             \\
               &                      &               & Bipartisan               & 2            & True, False                                                   \\
               &                      &               & Same State               & 2            & True, False                                                   \\
               &                      &               & Text (Title, Summary)    & 64           & act, administr, care, develop, ...                            \\ \cmidrule(lr){1-6}
Legislator     & 1084                 & 63            & Term                     & 2            & Representative, Senate                                        \\
               &                      &               & Party                    & 3            & Democrat, Independent, Republican                             \\
               &                      &               & State\&Territories       & 56           & AK, AL, AR, ...                                               \\
               &                      &               & Gender                   & 2            & Female, Male                                                  \\ \cmidrule(lr){1-6}
Registrant     & 6589                 & 118           & State                    & 54           & AK, AL, AR, ...                                               \\
               &                      &               & Text (Name, Description) & 64           & advoc, council, govern,                                       \\ \cmidrule(lr){1-6}
Lobbyist       & 25043                & 22            & Ethnicity                & 13           & Asian/GreaterEastAsian/EastAsian,... \\
               &                      &               & Party                    & 3            & Democrat, Independent, Republican                             \\
               &                      &               & Gender                   & 6            & Female, Male, Mostly Female, Mostly Male,...      \\ \bottomrule
\end{tabular}
}
\end{table*}
\begin{table*}[hbpt]
\centering
\caption{Graph Node Configuration}
\label{tab:graph_node_configuration}
\resizebox{0.9\textwidth}{!}{
\begin{tabular}{@{}llccccccc@{}}
   & Graph Configuration                          & Interest Group & Bill & Sponsor & Legislator & Registrant & Lobbyist & Related Bill \\ \midrule
1  & Base Graph (Interest Group + Bill)                   & O      & O    &         &            &            &          &              \\
2  & + sponsor                                    & O      & O    & O       &            &            &          &              \\
3  & + legislator (sponsor + cosponsor)           & O      & O    & O       & O          &            &          &              \\
4  & + registrant                                 & O      & O    &         &            & O          &          &              \\
5  & + lobbyist                                   & O      & O    &         &            &            & O        &              \\
6  & + relatedBill                                & O      & O    &         &            &            &          & O            \\
7  & + legislator-registrant                      & O      & O    & O       & O          & O          &          &              \\
8  & + legislator-lobbyist                        & O      & O    & O       & O          &            & O        &              \\
9  & + legislator-relatedBill                     & O      & O    & O       & O          &            &          & O            \\
10 & + registrant-lobbyist                        & O      & O    &         &            & O          & O        &              \\
11 & + registrant-relatedBill                     & O      & O    &         &            & O          &          & O            \\
12 & + lobbyist-relatedBill                       & O      & O    &         &            &            & O        & O            \\
13 & + legislator-registrant-lobbyist             & O      & O    & O       & O          & O          & O        &              \\
14 & + legislator-registrant-relatedBill          & O      & O    & O       & O          & O          &          & O            \\
15 & + legislator-lobbyist-relatedBill            & O      & O    & O       & O          &            & O        & O            \\
16 & + registrant-lobbyist-relatedBill            & O      & O    &         &            & O          & O        & O            \\
17 & + legislator-registrant-lobbyist-relatedBill & O      & O    & O       & O          & O          & O        & O            \\ \bottomrule
\end{tabular}
}
\end{table*}
\begin{table}
\centering
\caption{Graph Edge Size}
\label{tab:graph_edge_size}
\resizebox{0.9\linewidth}{!}{
\begin{tabular}{crrrr}
Graph Edges  & Client & Bill    & Legislator & Registrant \\ \midrule
Client       & -      & -       & -          & -          \\
Bill         & 602952 & 43712   & -          & -          \\
Legislator   & -      & 863369  & 341880     & -          \\
Sponsor Only & -      & 57466   & -          & -          \\
Registrant   & 29123  & 504056  & -          & -          \\
Lobbyist     & 107457 & 2062501 & 5611       & 31873      \\ \bottomrule
\end{tabular}
}
\end{table}

\subsection*{Entity Relations}
 From the overall graph illustrated in Fig.~\ref{fig:graph_structure}, we derive various graph configurations, as shown in Table~\ref{tab:graph_node_configuration}. The first row represents the Base Graph, which includes only Interest Group and Bill nodes along with edges between them indicating the groups lobbying on the bill. The second row, labeled `Base Graph + sponsor', adds legislator nodes who sponsor the bills and creates edges between bills and legislators. By adding new nodes and the corresponding new relations, we construct a total of 17 different graphs. We aim to identify the best graph for our bill position edge prediction task from these 17 graphs. The number of edges connecting nodes in the entire graph is provided in Table~\ref{tab:graph_edge_size}.

\subsection*{Parameter Search}
All input graphs use the fixed hyperparameters for model configuration listed in Table~\ref{tab:param_fix}, while the remaining GNN hyperparameters are optimized through a parameter search based on the values in Table~\ref{tab:param_search}. We run all hyperparameter combinations 1 time, and Table~\ref{tab:best_param} shows the best hyperparameter setting for each graph configuration. 


\begin{table}
\centering
\begin{minipage}{0.45\textwidth}
\centering
\caption{GNN Configuration}
\label{tab:param_fix}
\begin{tabular}{cc}
Parameter                       & Value                       \\ \midrule
Graph Model                     & SAGE                        \\ 
Number of Classes               & 3                           \\ 
Hidden Channels                 & 90                          \\ 
Weight Decay                    & 0.0                         \\ 
Activation Function             & ELU                         \\ 
Dropout Rate                    & 0.2                         \\ 
Decoder Predictor Type          & Concat                      \\ 
Decoder Layers                  & 2                           \\\bottomrule
\end{tabular}
\end{minipage}
\begin{minipage}{0.45\textwidth}
\centering
\caption{Parameter Search for GNN}
\label{tab:param_search}
\begin{tabular}{cc}
Parameter                & Value                \\ \midrule
num\_bases               & 3, 4                 \\
use\_skip                & True, False          \\
use\_decoder\_bn         & True, False          \\
use\_bn                  & True, False          \\
num\_layers              & 2, 3                 \\
learning rate            & 0.0005, 0.001, 0.005 \\ \bottomrule
\end{tabular}
\end{minipage}
\end{table}


\begin{table*}[t]
\centering
\caption{Best Parameter for each Graph Configuration}
\label{tab:best_param}
\resizebox{0.8\textwidth}{!}{
\begin{tabular}{@{}llcccccc@{}}
   & Graph Configuration                          & num\_layers & num\_bases & lr    & use\_bn & use\_skip & use\_decoder\_bn \\ \midrule
1  & Base Graph (Client + Bill)                   & 3           & 4          & 0.005 & FALSE   & TRUE      & TRUE             \\
2  & + sponsor                                    & 3           & 4          & 0.005 & FALSE   & TRUE      & FALSE            \\
3  & + legislator (sponsor + cosponsor)           & 3           & 3          & 0.005 & TRUE    & FALSE     & FALSE            \\
4  & + registrant                                 & 3           & 4          & 0.003 & FALSE   & FALSE     & TRUE             \\
5  & + lobbyist                                   & 3           & 3          & 0.005 & FALSE   & FALSE     & TRUE             \\
6  & + relatedBill                                & 3           & 3          & 0.005 & FALSE   & TRUE      & FALSE            \\
7  & + legislator-registrant                      & 3           & 4          & 0.005 & FALSE   & TRUE      & FALSE            \\
8  & + legislator-lobbyist                        & 3           & 3          & 0.005 & FALSE   & FALSE     & TRUE             \\
9  & + legislator-relatedBill                     & 3           & 4          & 0.003 & FALSE   & TRUE      & FALSE            \\
10 & + registrant-lobbyist                        & 2           & 4          & 0.005 & TRUE    & FALSE     & TRUE             \\
11 & + registrant-relatedBill                     & 3           & 4          & 0.005 & FALSE   & TRUE      & TRUE             \\
12 & + lobbyist-relatedBill                       & 3           & 4          & 0.005 & TRUE    & TRUE      & FALSE            \\
13 & + legislator-registrant-lobbyist             & 3           & 3          & 0.005 & FALSE   & FALSE     & FALSE            \\
14 & + legislator-registrant-relatedBill          & 3           & 4          & 0.005 & FALSE   & TRUE      & FALSE            \\
15 & + legislator-lobbyist-relatedBill            & 3           & 4          & 0.003 & FALSE   & FALSE     & FALSE            \\
16 & + registrant-lobbyist-relatedBill            & 2           & 3          & 0.003 & TRUE    & TRUE      & TRUE             \\
17 & + legislator-registrant-lobbyist-relatedBill & 3           & 4          & 0.003 & FALSE   & TRUE      & FALSE            \\ \bottomrule
\end{tabular}
}
\end{table*}

\begin{table*}
\centering
\caption{Best Hyperparameter Result for each Configuration}
\label{tab:best_param_result}
\resizebox{0.8\textwidth}{!}{
\begin{tabular}{@{}llccccc@{}}
   & Graph Configuration                          & Accuracy (Overall) & F1 Score (Overall) & F1 Score (Support) & F1 Score (Oppose) & F1 Score (Engage) \\ \midrule
1  & Base Graph (Client + Bill)                   & 76.07 ± 0.122      & 71.70 ± 0.209      & 77.04 ± 0.001      & 59.46 ± 0.007     & 78.60 ± 0.001     \\
2  & + sponsor                                    & 76.82 ± 0.037      & 72.31 ± 0.123      & 77.70 ± 0.001      & 60.06 ± 0.006     & 79.15 ± 0.000     \\
3  & + legislator (sponsor + cosponsor)           & 77.46 ± 0.067      & 73.41 ± 0.119      & 78.37 ± 0.001      & 62.29 ± 0.005     & 79.57 ± 0.001     \\
4  & + registrant                                 & 76.22 ± 0.303      & 71.14 ± 0.476      & 76.32 ± 0.004      & 57.65 ± 0.014     & 79.44 ± 0.003     \\
5  & + lobbyist                                   & 77.15 ± 0.134      & 72.14 ± 0.212      & 77.87 ± 0.002      & 58.56 ± 0.009     & 79.99 ± 0.000     \\
6  & + relatedBill                                & 76.30 ± 0.037      & 71.60 ± 0.083      & 77.14 ± 0.001      & 58.64 ± 0.004     & 79.03 ± 0.000     \\
7  & + legislator-registrant                      & 78.00 ± 0.058      & 73.97 ± 0.146      & 78.84 ± 0.000      & 62.97 ± 0.005     & 80.09 ± 0.001     \\
8  & + legislator-lobbyist                        & 78.51 ± 0.035      & 74.65 ± 0.087      & 79.19 ± 0.000      & 64.06 ± 0.006     & 80.70 ± 0.001     \\
9  & + legislator-relatedBill                     & 77.49 ± 0.047      & 73.64 ± 0.055      & 78.48 ± 0.001      & 63.18 ± 0.002     & 79.28 ± 0.001     \\
10 & + registrant-lobbyist                        & 76.54 ± 0.056      & 71.88 ± 0.291      & 77.02 ± 0.002      & 59.08 ± 0.019     & 79.53 ± 0.000     \\
11 & + registrant-relatedBill                     & 76.60 ± 0.027      & 71.69 ± 0.027      & 77.18 ± 0.002      & 58.33 ± 0.000     & 79.54 ± 0.000     \\
12 & + lobbyist-relatedBill                       & 77.63 ± 0.082      & 72.97 ± 0.139      & 78.12 ± 0.001      & 60.36 ± 0.004     & 80.44 ± 0.001     \\
13 & + legislator-registrant-lobbyist             & 78.16 ± 0.174      & 74.02 ± 0.130      & 79.01 ± 0.003      & 62.76 ± 0.001     & 80.29 ± 0.001     \\
14 & + legislator-registrant-relatedBill          & 77.95 ± 0.036      & 73.87 ± 0.152      & 78.67 ± 0.001      & 62.79 ± 0.009     & 80.15 ± 0.000     \\
15 & + legislator-lobbyist-relatedBill            & 78.52 ± 0.131      & 74.19 ± 0.143      & 79.44 ± 0.002      & 62.44 ± 0.002     & 80.68 ± 0.001     \\
16 & + registrant-lobbyist-relatedBill            & 76.95 ± 0.039      & 72.62 ± 0.050      & 77.54 ± 0.002      & 60.63 ± 0.004     & 79.70 ± 0.001     \\
17 & + legislator-registrant-lobbyist-relatedBill & 78.32 ± 0.032      & 74.21 ± 0.051      & 79.07 ± 0.000      & 63.11 ± 0.002     & 80.45 ± 0.000     \\ \bottomrule
\end{tabular}
}
\end{table*}

\subsection*{Best Parameter Result}
For the optimal setting for each case, we run the model five times, with the results shown in Table~\ref{tab:best_param_result}. Comparing the results, we observe that adding more nodes and relations generally leads to increased overall accuracy and F1 scores. Regarding the bill position classes, the \textit{Support} and \textit{Others} classes reach F1 scores in the high 70s, while \textit{Oppose} shows an F1 score in the 60s. Given that the \textit{Oppose} class has a lower proportion in the data, this suggests that predicting it is a more challenging task for the GNN. To select the best GNN model, we choose the eighth model, `Base Graph + legislator-lobbyist,' which has the highest overall F1 score of 74.65. This model also achieves the highest F1 score of 64.06 for the \textit{Oppose} class.

\section{Analysis1: Analyzing Bill Positions at Different Stages}
\label{sup:analysis1}


\begin{figure}
\centering
\includegraphics[width=\linewidth]{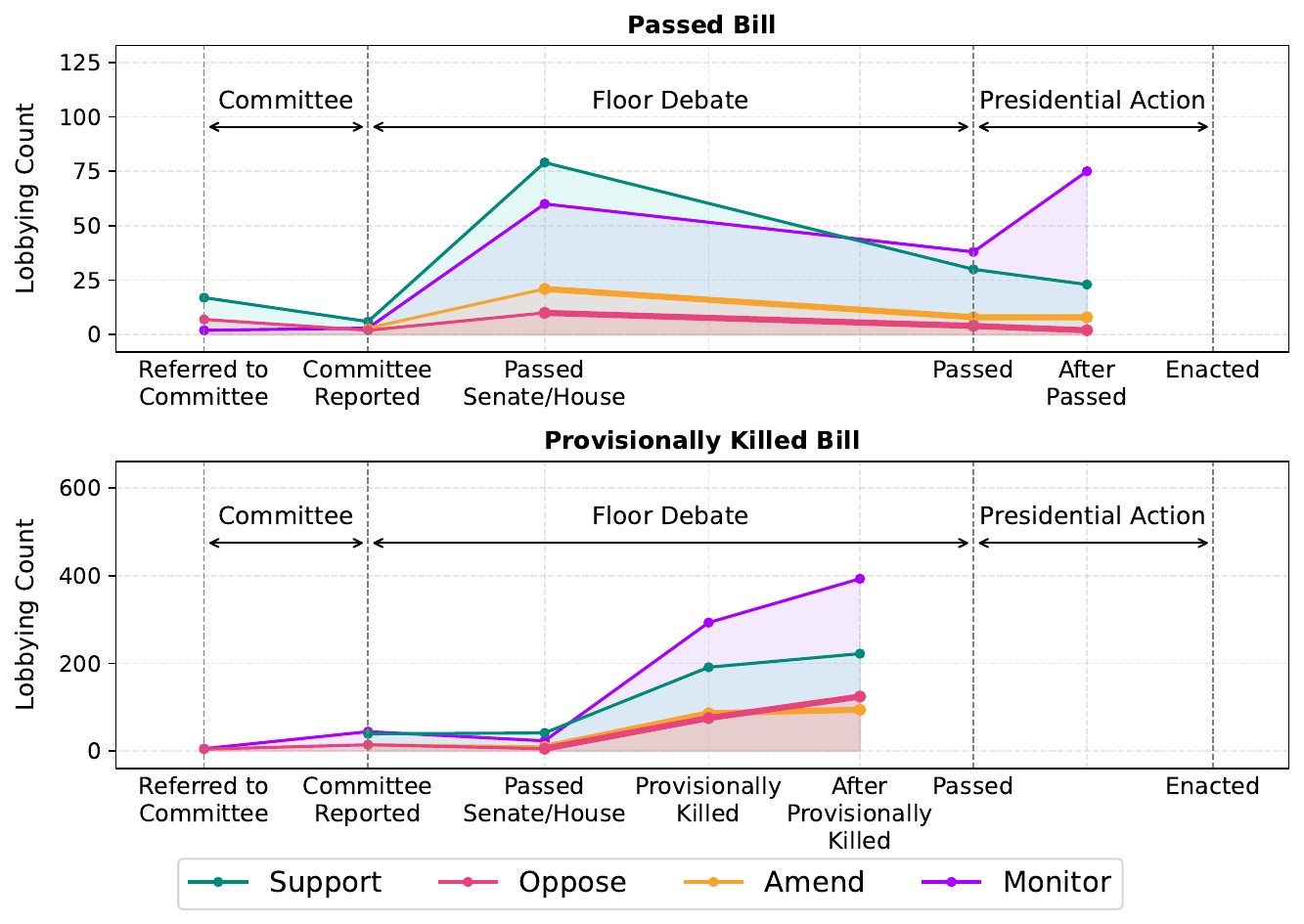}
\caption{Frequency of interest group positions across different legislative stages in `Passed' and `Provisionally Killed' (bills that failed to pass during floor debate) bills. For `Passed' bills, there is little change in the distribution of \textit{Oppose} and \textit{Amend}. However, for `Provisionally Killed' bills, the frequencies of \textit{Oppose} and \textit{Amend} rise sharply after passing one chamber (e.g., `Passed Senate/House').}
\label{fig:bill_stage_pass_prov_kill}
\end{figure}
\subsection*{Input Data Statistics for Bill Stage Analysis}
Table~\ref{tab:bill_stage_overall} presents the number of bills at each final status, along with the statistics of the interest groups that lobbied on these bills. These statistics are used for analyzing the bill positions of interest groups at the final stages, including `Enacted' and `Vetoed' bills. In addition to analyzing the patterns of `Enacted' and `Vetoed' bills, we also observe similar patterns in `Passed' and `Provisionally Killed' bills (those that failed to progress on the floor but were not passed).

\begin{table}
\centering
\caption{Statistics of Interest Groups and Bills at Final Bill Status}
\label{tab:bill_stage_overall}
\begin{tabular}{crr}
Bill Final Status    & \multicolumn{1}{c}{Bill} & \multicolumn{1}{c}{Interest Group} \\ \midrule
Enacted              & 715                      & 3640                               \\
Vetoed               & 3                        & 162                                \\
Passed               & 28                       & 128                                \\
Provisionally Killed & 74                       & 473                                \\ \bottomrule
\end{tabular}
\end{table}

\subsection*{Analysis of Interest Group Bill Positions Across Different Bill Stages}
Table~\ref{tab:bill_stage_enacted} and Table~\ref{tab:bill_stage_vetoed} show the number of bill positions lobbied at each bill stage for `Enacted' and `Vetoed' bills, respectively. Additionally, Tables~\ref{tab:bill_stage_passed} and~\ref{tab:bill_stage_prov_kill} present the number of bill positions lobbied at each stage for bills that reached the final stage as `Passed' and `Provisionally Killed'. Similar patterns are observed in both `Enacted' and `Vetoed' bills, as well as in the `Passed' and `Provisionally Killed' bills. The bill stage line plots in Fig.~\ref{fig:bill_stage_pass_prov_kill} illustrate that `Passed' bills consistently show low proportions of \textit{Oppose} and \textit{Amend}, with higher proportions of \textit{Support} and \textit{Monitor}. In contrast, `Provisionally Killed' bills show an increase in \textit{Oppose} and \textit{Amend}. These findings suggest a correlation between the number of bill positions and whether a bill progresses to the next major stage across different bill statuses.

\begin{table}[t]
\centering
\begin{minipage}{0.45\textwidth}
\centering
\caption{Enacted Bill}
\label{tab:bill_stage_enacted}
\resizebox{\linewidth}{!}{
\begin{tabular}{crrrr|r}
Bill Stage            & Support & Oppose & Amend & Monitor & Sum   \\ \midrule
Referred to Committee & 337     & 18     & 40    & 140     & 535   \\
Committee Reported    & 253     & 22     & 64    & 206     & 545   \\
Passed Senate/House   & 1193    & 118    & 378   & 984     & 2673  \\
Passed                & 345     & 43     & 112   & 229     & 729   \\
Enacted               & 3716    & 505    & 925   & 3708    & 8854  \\
After Enacted         & 3940    & 617    & 1156  & 8718    & 14431 \\ \midrule
Sum                   & 10449   & 1495   & 2956  & 14863   & 29763 \\ \bottomrule
\end{tabular}
}
\end{minipage}
\begin{minipage}{0.45\textwidth}
\centering
\caption{Vetoed Bill}
\label{tab:bill_stage_vetoed}
\resizebox{\linewidth}{!}{
\begin{tabular}{crrrr|r}
Bill Stage            & Support & Oppose & Amend & Monitor & Sum \\ \midrule
Referred to Committee & 2       & 0      & 0     & 5       & 7   \\
Committee Reported    & 14      & 4      & 9     & 21      & 48  \\
Passed Senate/House   & 21      & 3      & 11    & 25      & 60  \\
Vetoed                & 26      & 13     & 48    & 28      & 115 \\
After Vetoed          & 31      & 18     & 30    & 31      & 110 \\ \midrule
Sum                   & 102     & 42     & 101   & 119     & 364 \\ \bottomrule
\end{tabular}
}
\end{minipage}
\begin{minipage}{0.45\textwidth}
\centering
\caption{Passed Bill}
\label{tab:bill_stage_passed}
\resizebox{\linewidth}{!}{
\begin{tabular}{crrrr|r}
Bill Stage            & Support & Oppose & Amend & Monitor & Sum \\ \midrule
Referred to Committee & 17      & 7      & 0     & 2       & 26  \\
Committee Reported    & 6       & 2      & 3     & 3       & 14  \\
Passed Senate/House   & 35      & 1      & 12    & 16      & 64  \\
Passed                & 30      & 4      & 8     & 38      & 80  \\
After Passed          & 23      & 2      & 8     & 75      & 108 \\ \midrule
Sum                   & 111     & 16     & 31    & 134     & 292 \\ \bottomrule
\end{tabular}
}
\end{minipage}
\begin{minipage}{0.45\textwidth}
\centering
\caption{Provisionally Killed Bill}
\label{tab:bill_stage_prov_kill}
\resizebox{\linewidth}{!}{
\begin{tabular}{crrrr|r}
Bill Stage                 & Support & Oppose & Amend & Monitor & Sum  \\ \midrule
Referred to Committee      & 0       & 4      & 0     & 2       & 6    \\
Committee Reported         & 39      & 14     & 14    & 44      & 111  \\
Passed Senate/House        & 41      & 5      & 9     & 23      & 78   \\
Temporarily Rejected       & 191     & 75     & 86    & 293     & 645  \\
After Temporarily Rejected & 217     & 112    & 83    & 378     & 790  \\ \midrule
Sum                        & 488     & 210    & 192   & 740     & 1630 \\ \bottomrule
\end{tabular}
}
\end{minipage}
\end{table}

\section{Analysis 2: Bill Position for Firm Size Analysis}
\label{sup:analysis2}

\subsection*{Input Data Statistics for Firm Size Analysis}

Table~\ref{tab:firm_size_summary} presents the summary statistics of the firm-year dataset used for the firm size analysis. The dataset spans from 2009 to 2022, with observations ranging from 5,421 to 7,014 firms per year. For each year, the table reports the average proportions of different lobbying outcomes: whether firms lobbied at all (Lobby), and the specific lobbying positions employed (Support, Oppose, Amend, and Monitor). Standard deviations for each measure are presented in parentheses. The table also includes the log-transformed employment values, which serve as the key explanatory variable in our analysis of how firm size relates to lobbying behavior.

\begin{table*}
\centering
\caption{Data Statistics for Firm Size Analysis}
\label{tab:firm_size_summary}
\resizebox{0.8\textwidth}{!}{
\begin{tabular}{cccccccc}
 Year &  Obs. &         Lobby &       Support &        Oppose &         Amend &       Monitor & Log Employment \\
\midrule
 2009 &  7014 & 0.086 (0.280) & 0.198 (0.343) & 0.036 (0.159) & 0.246 (0.362) & 0.519 (0.434) &  6.315 (2.618) \\
 2010 &  6943 & 0.086 (0.280) & 0.207 (0.343) & 0.042 (0.174) & 0.236 (0.365) & 0.515 (0.435) &  6.309 (2.654) \\
 2011 &  6931 & 0.074 (0.261) & 0.252 (0.348) & 0.037 (0.147) & 0.201 (0.326) & 0.510 (0.419) &  6.300 (2.698) \\
 2012 &  7000 & 0.077 (0.267) & 0.271 (0.381) & 0.038 (0.165) & 0.162 (0.312) & 0.529 (0.435) &  6.245 (2.748) \\
 2013 &  6912 & 0.071 (0.258) & 0.245 (0.370) & 0.041 (0.160) & 0.154 (0.310) & 0.560 (0.433) &  6.262 (2.766) \\
 2014 &  6748 & 0.077 (0.267) & 0.219 (0.357) & 0.028 (0.118) & 0.115 (0.272) & 0.637 (0.422) &  6.300 (2.772) \\
 2015 &  6575 & 0.073 (0.261) & 0.241 (0.353) & 0.025 (0.113) & 0.153 (0.306) & 0.582 (0.420) &  6.318 (2.782) \\
 2016 &  6451 & 0.077 (0.266) & 0.296 (0.390) & 0.023 (0.113) & 0.142 (0.283) & 0.539 (0.428) &  6.338 (2.791) \\
 2017 &  6332 & 0.072 (0.259) & 0.245 (0.366) & 0.027 (0.132) & 0.165 (0.313) & 0.563 (0.433) &  6.357 (2.783) \\
 2018 &  6239 & 0.075 (0.263) & 0.286 (0.411) & 0.020 (0.115) & 0.134 (0.282) & 0.560 (0.445) &  6.417 (2.766) \\
 2019 &  6291 & 0.070 (0.256) & 0.293 (0.413) & 0.031 (0.158) & 0.155 (0.320) & 0.521 (0.446) &  6.393 (2.767) \\
 2020 &  6162 & 0.092 (0.290) & 0.247 (0.386) & 0.020 (0.107) & 0.126 (0.297) & 0.606 (0.444) &  6.398 (2.759) \\
 2021 &  5886 & 0.099 (0.299) & 0.219 (0.375) & 0.013 (0.097) & 0.107 (0.284) & 0.661 (0.435) &  6.508 (2.740) \\
 2022 &  5421 & 0.046 (0.211) & 0.164 (0.329) & 0.007 (0.056) & 0.045 (0.180) & 0.785 (0.361) &  6.644 (2.714) \\
\midrule
\multicolumn{8}{l}{\footnotesize Note: Values represent sample means; standard deviations are reported in parentheses.}
\end{tabular}
}
\end{table*}

\subsection*{Alternative Specifications for Firm Size and Lobbying Pattern Analysis}

While Fig.~\ref{fig:firm_size_prob} presents the logit and Dirichlet regression results with year fixed effects, we further explore alternative model specifications to examine how the regression results change as we add or remove additional controls. These controls include industry fixed effects (using the first two digits of NAICS code), log of net property, plant, and equipment (PPENT), and log of cost of goods sold (COGS). Both PPENT and COGS, along with the employment variable, are from the Compustat database. 

Table~\ref{tab:SI_dirichlet_qoi} and Table~\ref{tab:logistic_qoi} present the same quantity of interest as shown in Fig.~\ref{fig:firm_size_prob}: the average difference in predicted probabilities of lobbying outcomes between firms at the 90th and 10th percentile of employment. In Table~\ref{tab:SI_dirichlet_qoi}, the positive and significant effect of firm size on adopting a monitoring position is consistent across all model specifications. In addition, the negative effect of firm size on opposing positions is statistically significant in all but one model specification. Table~\ref{tab:logistic_qoi} demonstrates that the positive effect of firm size on the decision to lobby remains consistent across all model specifications. These results support our main findings from Fig.~\ref{fig:firm_size_prob}: larger firms not only lobby more frequently but also tend to adopt monitoring positions more often, while smaller firms are more likely to take reactive and explicit positions.

\begin{table*}
\centering
\caption{Dirichlet Regression Results: Effect of Firm Size on Predicted Probability of Lobbying Position}

\resizebox{\textwidth}{!}{%
\begin{tabular}{lccccc}
 & No Fixed Effects & Year FE & Year + NAICS FE & Year FE + Log PPENT + Log COGS & Year + NAICS FE + Log PPENT + Log COGS \\
\midrule
Support & -0.014 (0.011) & -0.015 (0.012) & -0.024 (0.013)$^*$ & -0.027 (0.019) & -0.043 (0.024)$^*$ \\
Oppose  & -0.017 (0.005)$^{***}$ & -0.018 (0.005)$^{***}$ & -0.025 (0.007)$^{***}$ & -0.009 (0.009) & -0.020 (0.011)$^*$ \\
Amend   & -0.014 (0.007)$^*$ & -0.013 (0.007)$^*$ & -0.011 (0.009) & -0.042 (0.014)$^{***}$ & -0.014 (0.017) \\
Monitor & 0.045 (0.020)$^{**}$ & 0.045 (0.020)$^{**}$ & 0.060 (0.023)$^{**}$ & 0.078 (0.032)$^{**}$ & 0.077 (0.042)$^*$ \\
\midrule
\multicolumn{6}{l}{\footnotesize Note: Clustered standard errors are in parentheses. *p$<$0.1; **p$<$0.05; ***p$<$0.01.} \\
\end{tabular}%
}
\label{tab:SI_dirichlet_qoi}
\end{table*}

\begin{table*}
\centering
\caption{Logistic Regression Results: Effect of Firm Size on Predicted Probability of Lobbying}
\resizebox{\textwidth}{!}{%
\begin{tabular}{lccccc}
 & No Fixed Effects & Year FE & Year + NAICS FE & Year FE + Log PPENT + Log COGS & Year + NAICS FE + Log PPENT + Log COGS \\
\midrule
Lobby & 0.193 (0.008)$^{***}$ & 0.195 (0.008)$^{***}$ & 0.205 (0.009)$^{***}$ & 0.053 (0.012)$^{***}$ & 0.055 (0.015)$^{***}$ \\
\midrule
\multicolumn{6}{l}{\footnotesize Note: Clustered standard errors are in parentheses. *p$<$0.1; **p$<$0.05; ***p$<$0.01.} \\\\
\end{tabular}%
}

\label{tab:logistic_qoi}
\end{table*}

\section{Analysis 3: How Bill Positions Differ Across Subjects}
\label{sup:analysis3}
\begin{table*}
\centering
\caption{Bill Position Ratio across Bill Subjects}
\label{tab:bill_subject_position_ratio}
\begin{tabular}{@{}crcccc@{}}
Bill Subject                        & Unique Bills & Avg. Lobbying per Bill & Support Ratio & Oppose Ratio & Engage Ratio \\ \midrule
Taxation                            & 239          & 27.590                 & 0.833         & 0.043        & 0.124        \\
Health                              & 342          & 24.395                 & 0.789         & 0.084        & 0.127        \\
Transportation and public works     & 83           & 38.880                 & 0.745         & 0.073        & 0.182        \\
Commerce                            & 74           & 27.757                 & 0.725         & 0.121        & 0.154        \\
Science, technology, communications & 51           & 28.255                 & 0.725         & 0.079        & 0.196        \\
Crime and law enforcement           & 86           & 23.279                 & 0.724         & 0.141        & 0.134        \\
Immigration                         & 57           & 23.211                 & 0.689         & 0.213        & 0.098        \\
Energy                              & 72           & 24.764                 & 0.661         & 0.161        & 0.178        \\
Labor and employment                & 130          & 23.892                 & 0.660         & 0.220        & 0.120        \\
Government operations and politics  & 129          & 26.899                 & 0.630         & 0.252        & 0.118        \\
Finance and financial sector        & 89           & 23.225                 & 0.620         & 0.187        & 0.193        \\
Public lands and natural resources  & 51           & 15.843                 & 0.557         & 0.354        & 0.089        \\
Environmental protection            & 108          & 26.898                 & 0.555         & 0.280        & 0.165        \\
Armed forces and national security  & 78           & 44.577                 & 0.533         & 0.111        & 0.357        \\
Economics and public finance        & 162          & 73.691                 & 0.516         & 0.063        & 0.421        \\ \bottomrule
\end{tabular}
\end{table*}

We provide further details on the bills, highlighting the differences in bill position ratios based on their subject, with high ratios of \textit{Support}, \textit{Oppose}, and \textit{Engage}.

\subsection*{Bills with a high \textit{Support} ratio}
\begin{figure*}[ht]
\centering
\begin{minipage}{0.48\linewidth}
    \centering
    \includegraphics[width=\linewidth]{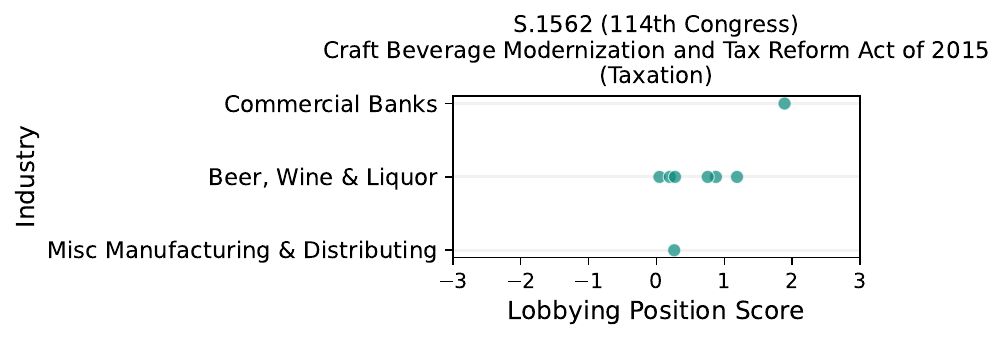} 
    \includegraphics[width=\linewidth]{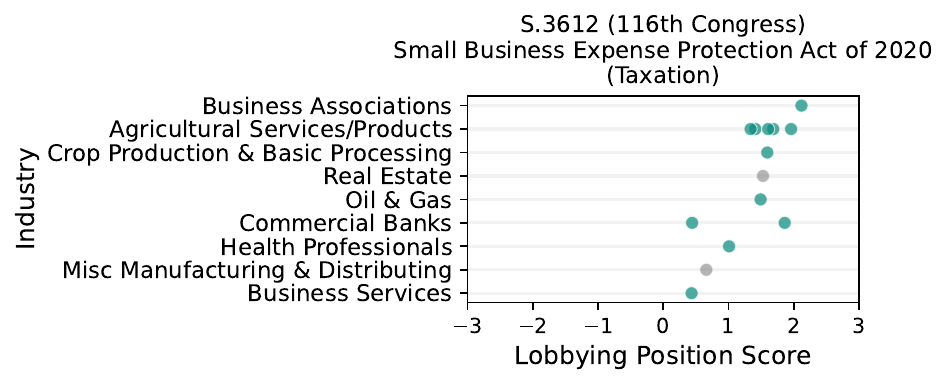}
\end{minipage}
\begin{minipage}{0.48\linewidth}
    \includegraphics[width=\linewidth]{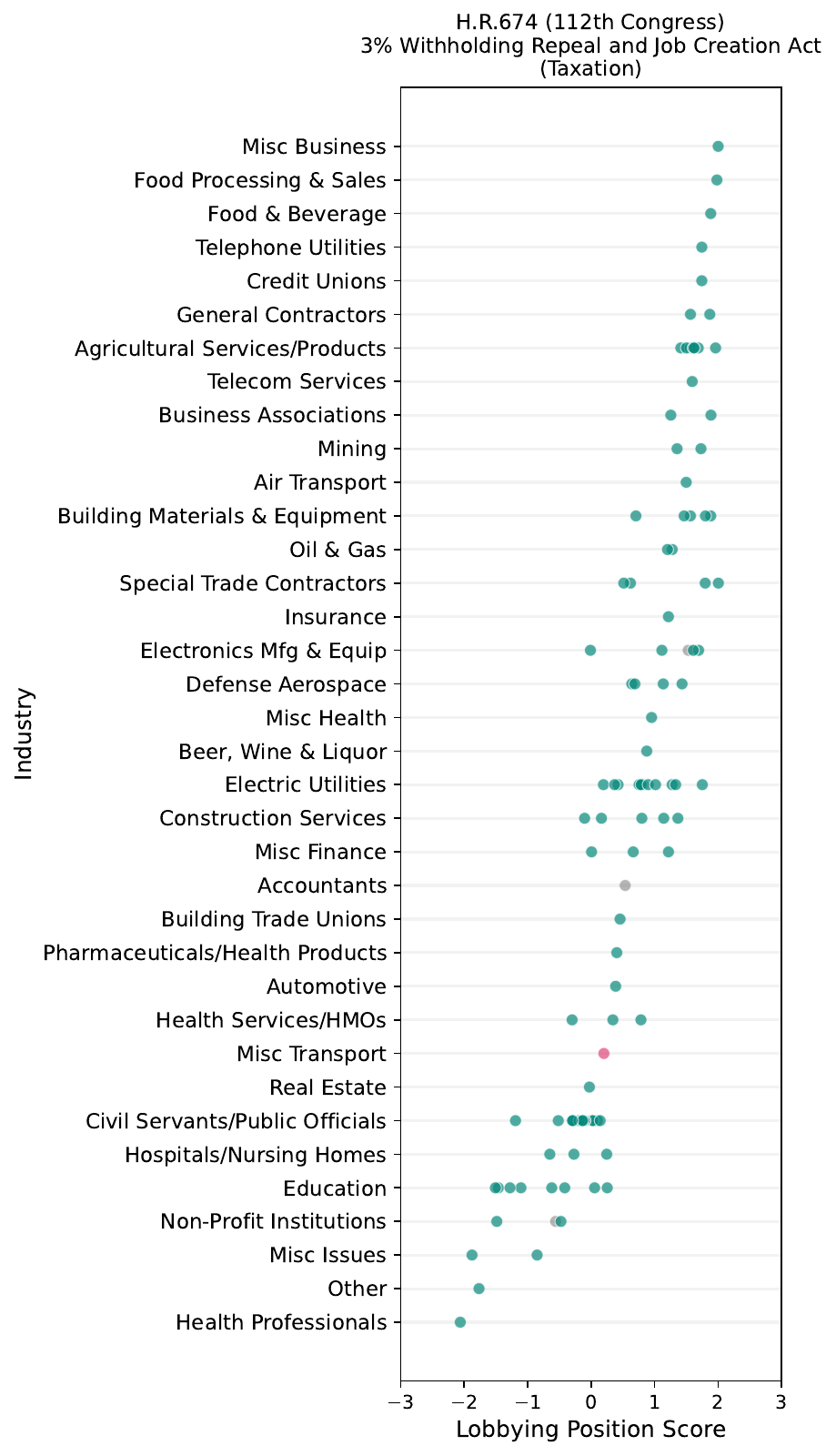}
\end{minipage}

\caption{Bills with high \textit{Support} ratios, illustrating bill positions across industries}
\label{fig:support_industry_bill_position}
\end{figure*}

We present additional details on the four bills with high \textit{Support} ratios discussed in the main paper, with Fig.~\ref{fig:support_industry_bill_position} illustrating these bills along with the distribution of bill positions by interest groups for each bill.

\paragraph{\texttt{S. 260 (115th) - Repeal of Independent Payment Advisory Board under the Affordable Care Act}} aims to reduce government intervention in healthcare by repealing the Independent Payment Advisory Board (IPAB) provisions of the Affordable Care Act. Support for this bill comes predominantly from medical and healthcare-related industries. Interest groups representing \textit{Health Professionals}, \textit{Hospitals/Nursing Homes}, \textit{Human Rights}, and \textit{Pharmaceuticals/Health Products} have actively lobbied in favor of this legislation, as shown in Fig.~\ref{fig:bill_industry_bill_position} (B).

\paragraph{\texttt{S. 1562 (114th) - Reform Taxation of Alcoholic Beverages}} addresses the reform of alcoholic beverage taxation. The support for this legislation comes primarily from industries related to alcohol production and distribution. As shown in Fig.~\ref{fig:support_industry_bill_position}, there is distinct lobbying from sectors such as \textit{Beer, Wine \& Liquor}, which directly benefit from the proposed tax reductions.

\paragraph{\texttt{S. 3612 (116th) - Small Business Expense Protection Act of 2020}} provides tax deductions for small businesses, with broad support from industries benefiting from these financial incentives. Various sectors, including \textit{Miscellaneous Manufacturing \& Distributing}, \textit{Health Professionals}, \textit{Commercial Banks}, \textit{Oil \& Gas}, \textit{Real Estate}, and \textit{Agricultural Services/Products}, are actively involved in lobbying for this legislation. The figure highlights how these industries have aligned interests in supporting the bill’s provisions.

\paragraph{\texttt{H.R. 674 (112th) - Veterans' Pensions and Compensation}} provides financial benefits to veterans, including exemptions from tax withholding and the introduction of veterans’ support programs. The broad scope of benefits has garnered support not only from veterans' groups but also from industries that serve them. These industries, ranging from healthcare to financial services, demonstrate how targeted benefits to a specific group, like veterans, can lead to involvement from multiple sectors. 

\subsection*{Bills with a high \textit{Oppose} ratio}
\begin{figure*}[ht]
\centering

\begin{minipage}{0.48\linewidth}
    \centering
    \includegraphics[width=\linewidth]{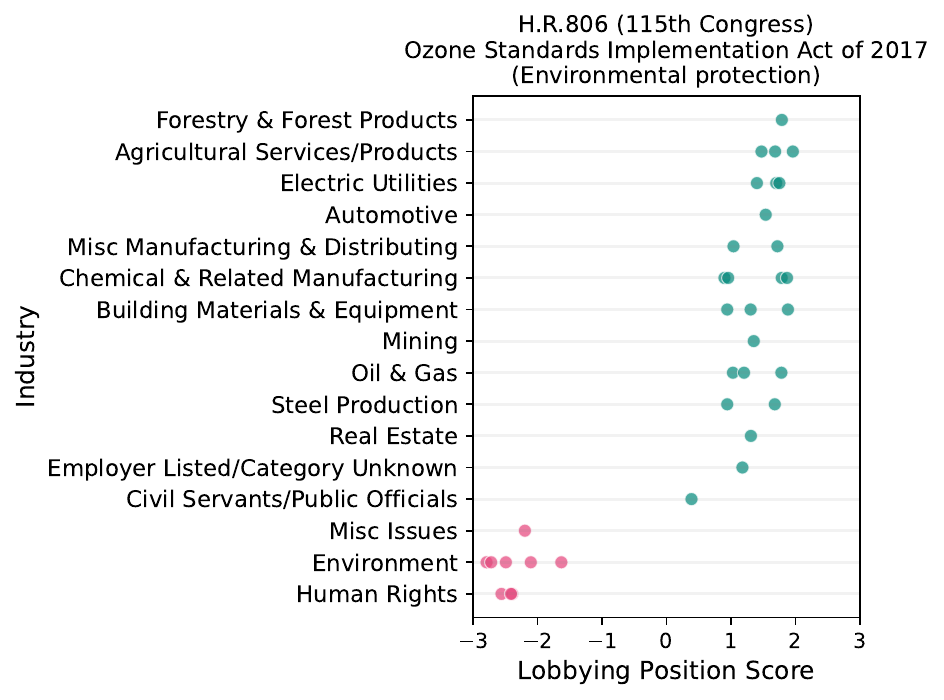}
\end{minipage}
\begin{minipage}{0.48\linewidth}
    \includegraphics[width=\linewidth]{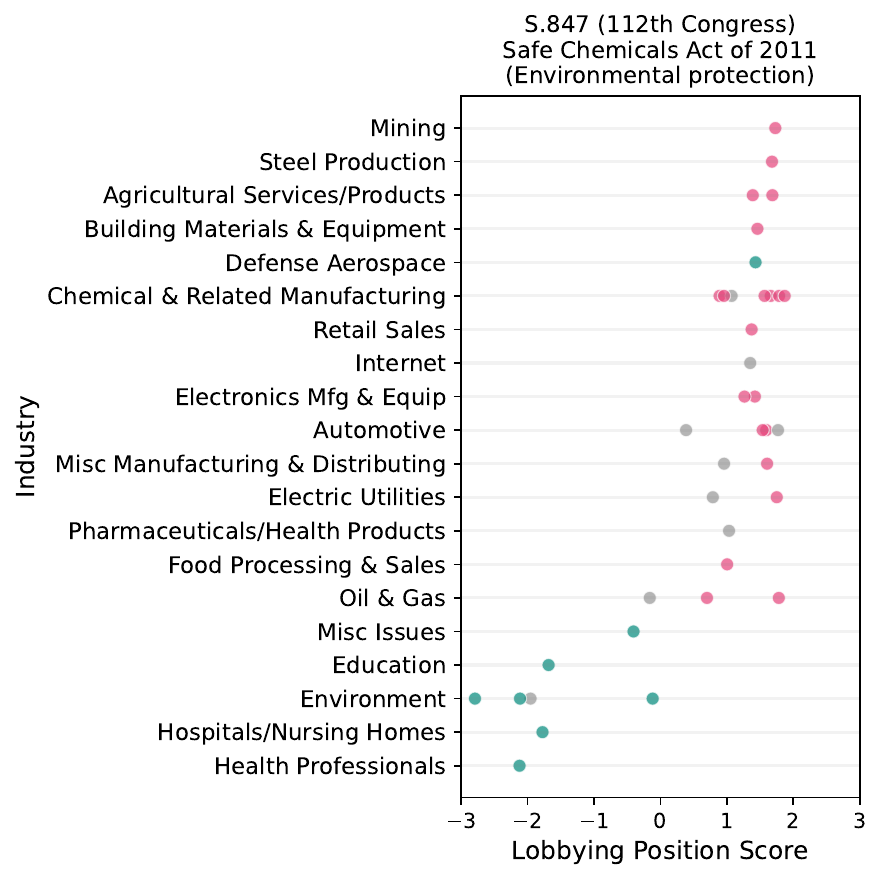}
\end{minipage}

\caption{Bills with high \textit{Oppose} ratios, illustrating bill positions across industries}
\label{fig:oppose_industry_bill_position}
\end{figure*}

For bills with high \textit{Oppose} ratios, we provide further details, with Fig.~\ref{fig:oppose_industry_bill_position}.

\paragraph{\texttt{H.R. 3409 (112th) - Stop the War on Coal Act of 2012}}
This bill aimed to deregulate the coal industry by limiting coal mining regulations, banning EPA greenhouse gas rules for climate change, and requiring transparent analysis of regulatory impacts. While the coal industry and commercial entities supported this bill, environmental groups strongly opposed it due to concerns about its potential harm to environmental protection and climate change mitigation efforts as shown in Fig.~\ref{fig:bill_industry_bill_position} (C).

\paragraph{\texttt{H.R. 806 (115th) - Ozone Standards Implementation Act of 2017}}
This bill delays the 2015 ozone standards implementation and extends the review period to 10 years. It allows for revisions based on feasibility and exempts certain regions from penalties. The broad involvement of various industries, including environmental and energy sectors, results in a high \textit{Engage} ratio, as diverse stakeholders engage in lobbying efforts to influence the bill's provisions.

\paragraph{\texttt{S. 847 (112th) - Safe Chemicals Act of 2011}}
The Safe Chemicals Act aimed to strengthen the management of chemicals, especially in protecting vulnerable populations and the environment, by requiring the safety assessment of commercial chemicals, reducing the use of harmful substances, and mandating transparency in health and environmental impacts. While environmental interest groups strongly supported the bill, commercial industries opposed it due to its potential impact on the chemical industry’s regulatory burden and operations.

\subsection*{Bills with a high \textit{Engage} ratio}
\begin{figure*}[ht]
\centering

\begin{minipage}{0.49\linewidth}
    \centering
    \includegraphics[width=\linewidth]{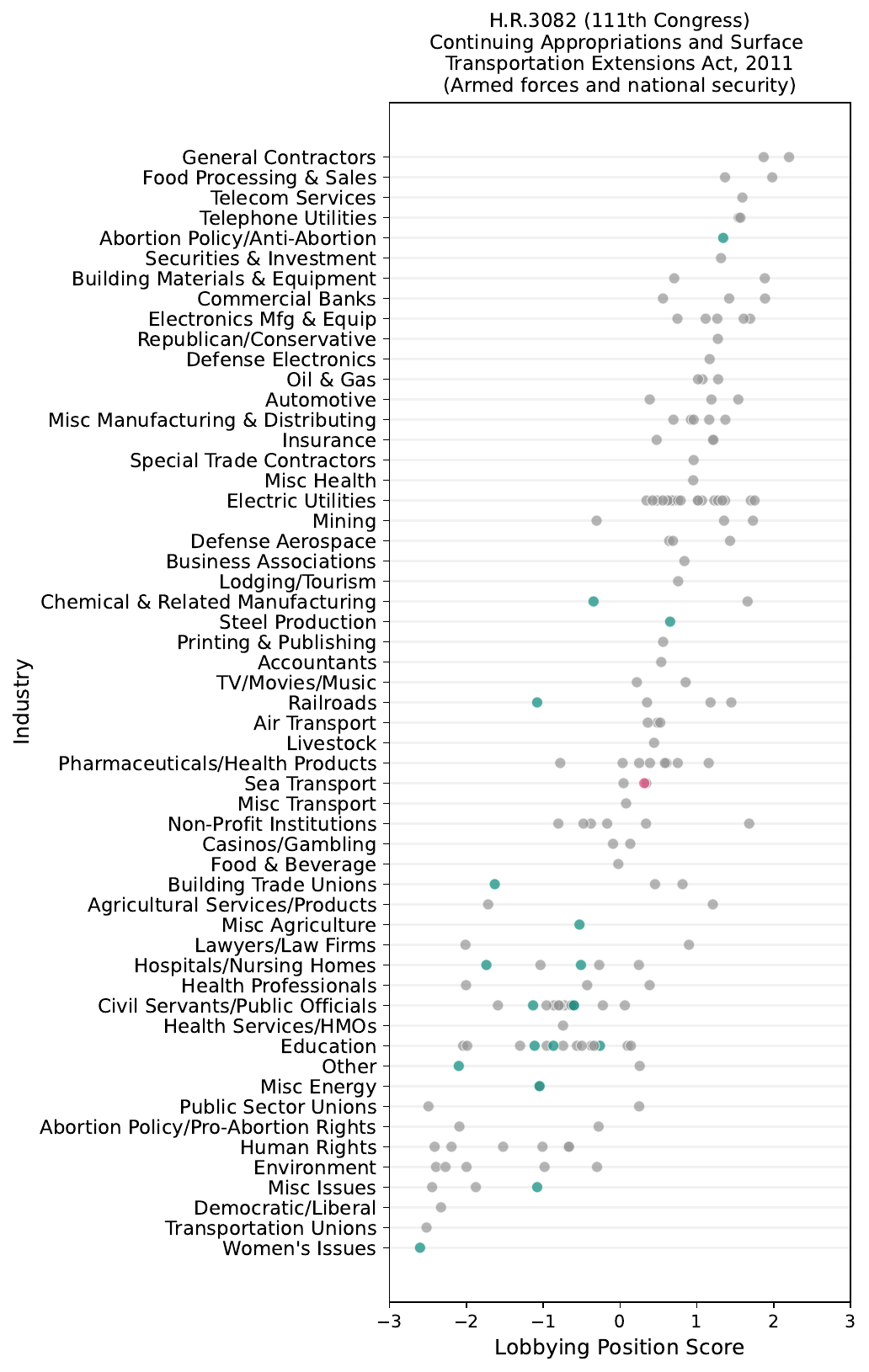}
\end{minipage}
\begin{minipage}{0.49\linewidth}    
    \includegraphics[width=\linewidth]{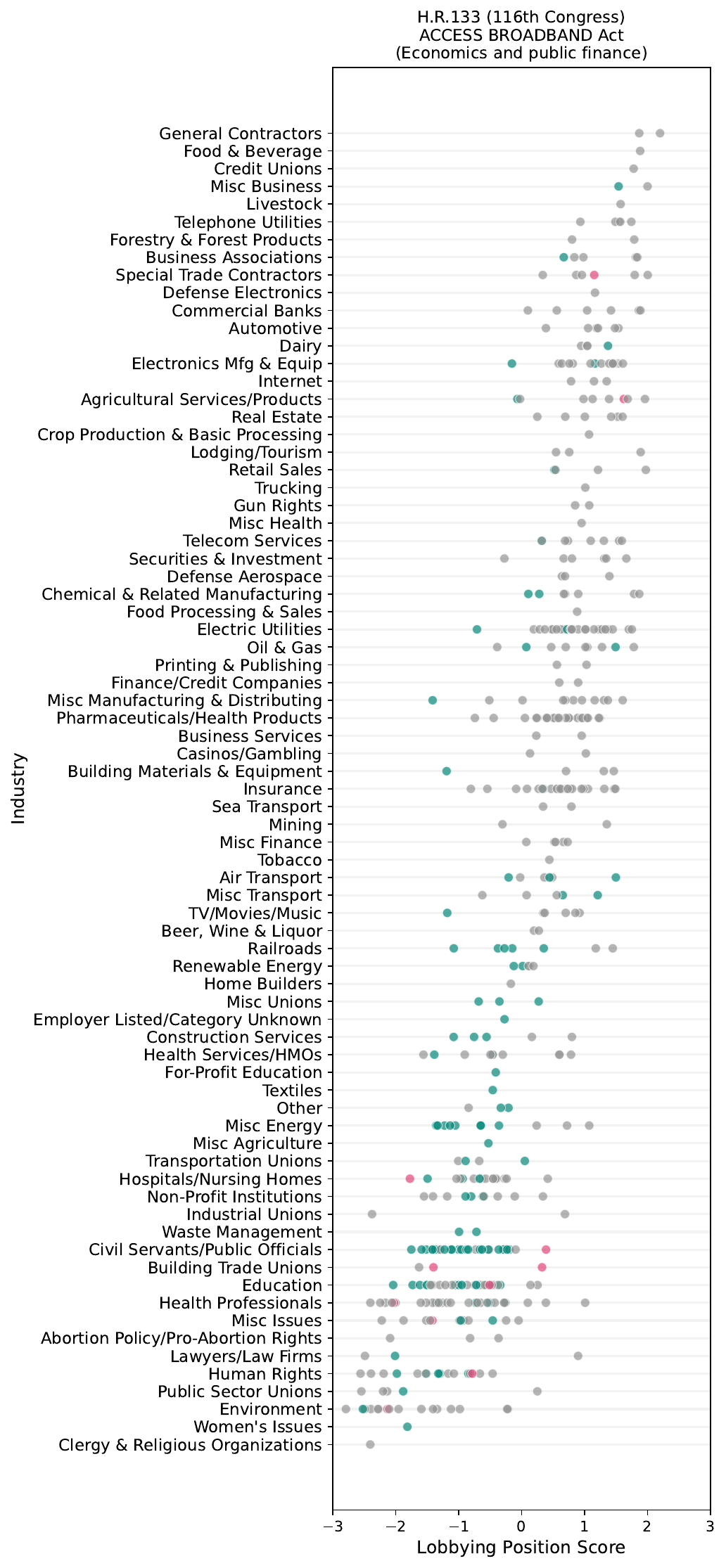}
\end{minipage}

\caption{Bills with high \textit{Engage} ratios, illustrating bill positions across industries}
\label{fig:engage_industry_bill_position}
\end{figure*}

Additional details on bills with high \textit{Engage} ratios are presented, with Fig.~\ref{fig:engage_industry_bill_position}.

\paragraph{\texttt{H.R. 8337 (116th) - Continuing Appropriations Act, 2021 and Other Extensions Act}}
This bill provides appropriations to federal agencies and extends expiring programs addressing health care, surface transportation, agriculture, veterans benefits, and other key areas. With its broad scope, it attracts wide support from various sectors, including healthcare, transportation, and defense industries. The bill's diverse allocation of funds leads to a high \textit{Engage} ratio, reflecting the wide range of stakeholders involved in lobbying efforts as shown in Fig.~\ref{fig:bill_industry_bill_position} (D).

\paragraph{\texttt{H.R. 3082 (111th) - Continuing Appropriations and Surface Transportation Extensions Act, 2011}}
This legislation involves budget allocations for sectors like defense, small businesses, and transportation. Its impact on multiple industries leads to broad lobbying participation, contributing to a high \textit{Engage} ratio due to the diverse interests involved.

\paragraph{\texttt{H.R. 133 (116th) - Consolidated Appropriations Act, 2021}}
This bill includes appropriations for rural development, agriculture, science, defense, energy, finance, and environmental protection. Due to its comprehensive funding provisions, it engages multiple industries, from defense to environmental protection, resulting in a high \textit{Engage} ratio as various sectors seek to influence policy and secure financial support for their interests.

\section{Analysis 4: How Interest Group Preferences Vary Across Industries}
\label{sup:analysis4}
\begin{figure*}
\centering
\includegraphics[width=0.5\textwidth]{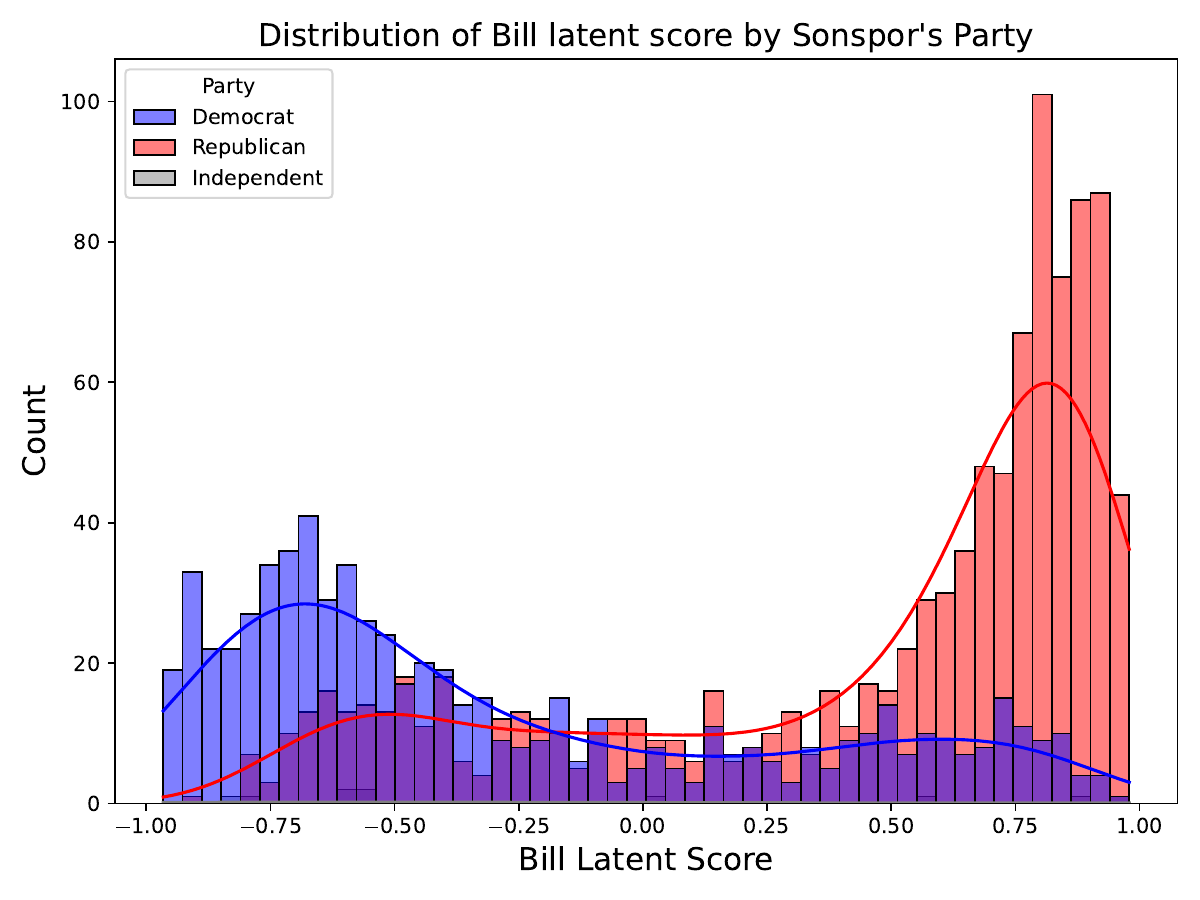}
\caption{Sponsor party's distribution of the bill's latent score.}
\label{fig:sponsor_party_distribution}
\end{figure*}

Fig.~\ref{fig:bill_industry_bill_position} (A) shows bill position distributions across bill and interest groups. In this section, we further explain the sponsor's party distribution with bill latent scores in Fig.~\ref{fig:sponsor_party_distribution}. This figure clearly illustrates the polarized distribution between parties, with a clear mapping to the latent scores, where greater distances in the latent scores correspond to affiliation with different parties.

\end{document}